\documentstyle[aps,prb]{revtex}
\draft

\begin{document}

\title{Finite-Size Scaling of the Ground State Parameters of the 
Two-Dimensional Heisenberg Model}

\author{Anders W. Sandvik}
\address{Department of Physics, University of Illinois at Urbana-Champaign,
1110 West Green Street, Urbana, Illinois 61801}

\date{July 10, 1997}
\maketitle{}

\begin{abstract}
The ground state parameters of the two-dimensional $S=1/2$ antiferromagnetic 
Heisenberg model are calculated using the Stochastic Series Expansion 
quantum Monte Carlo method for $L \times L$ lattices with $L$ up to $16$. 
The finite-size results for the energy $E$, the sublattice magnetization $M$,
the long-wavelength susceptibility $\chi_\perp (q=2\pi/L)$, and the spin 
stiffness $\rho_{\rm s}$, are extrapolated to the thermodynamic limit using 
fits to polynomials in $1/L$, constrained by scaling forms previously obtained
from renormalization group calculations for the nonlinear $\sigma$ model and 
chiral perturbation theory. The results are fully consistent with the 
predicted leading finite-size corrections and are of sufficient accuracy for 
extracting also subleading terms. The subleading energy correction 
($\sim 1/L^4$) agrees with chiral perturbation theory to within a statistical 
error of a few percent, thus providing the first numerical confirmation of the 
finite-size scaling forms to this order. The extrapolated ground state energy 
per spin, $E=-0.669437(5)$, is the most accurate estimate reported to date. 
The most accurate Green's function Monte Carlo (GFMC) result is slightly 
higher than this value, most likely due to a small systematic error 
originating from ``population control'' bias in GFMC. The other extrapolated 
parameters are $M=0.3070(3)$, $\rho_{\rm s} = 0.175(2)$, $\chi_\perp = 
0.0625(9)$, and the spinwave velocity $c=1.673(7)$. The statistical errors are
comparable with those of the best previous estimates, obtained by fitting loop
algorithm quantum Monte Carlo data to finite-temperature scaling forms. 
Both $M$ and $\rho_s$ obtained from the finite-$T$ data are, however, 
a few error bars higher than the present estimates. It is argued that the 
$T=0$ extrapolations performed here are less sensitive to effects of neglected 
higher-order corrections and therefore should be more reliable.
\end{abstract}


\section{Introduction}

In the nonlinear $\sigma$ model description of the two-dimensional (2D)
Heisenberg model,\cite{csh} the low-energy and low-temperature properties of 
the system are completely determined by three ground state parameters; the 
sublattice magnetization $M$, the spin stiffness constant $\rho_s$, and the 
spinwave velocity $c$. Their values are not given by the theory, however, 
but have to be determined starting from the microscopic Hamiltonian. 
A large number of calculations of the ground state parameters
have been carried out. The antiferromagnetically ordered ground state, 
which has been established rigorously only for $S > 1/2$,\cite{s1order} 
was first convincingly confirmed also for $S=1/2$ in a quantum Monte Carlo 
(QMC) study by Reger and Young.\cite{reger} The sublattice magnetization 
obtained this way, $M \approx 0.30$ (in units where the N\'eel state has 
$M=1/2$), also indicated that spinwave theory \cite{spinwave1,spinwave2} 
gives a surprisingly good quantitative description of the ground state. The 
same conclusion was reached by Singh,\cite{singh} who carried out a series 
expansion around the Ising limit, and found $M \approx 0.30$, $\rho_s \approx 
0.18J$, and $c \approx 1.7J$ ($J$ is the nearest-neighbor exchange coupling), 
all in good agreement with spinwave theory including the $1/S$ corrections.
\cite{spinwave2} Subsequent higher-order spinwave calculations showed that 
the $1/S^2$ corrections to $M$, $\rho_s$ and $c$ indeed are small.
\cite{hamer,igarashi,canali} Several other QMC simulations,
\cite{barnes,carlson,gross,trivedi,liang,runge1,runge2,sauer,wiese,beard}
exact diagonalizations,\cite{schulz1,schulz2,einarsson} 
as well as series expansions to higher orders,\cite{weihong}
have confirmed and improved on the accuracy of the above
estimates. The presently most accurate calculations
\cite{runge2,wiese,beard,weihong} indicate that the true values of 
the ground state parameters deviate from their $1/S^2$ spinwave values by 
only 1-2\% or less.

For most practical purposes (such as extracting $J$ for a system from 
experimental data), the ground state parameters of the 2D Heisenberg model
are now known to quite sufficient accuracy. However, there are still reasons 
to go to even higher precision. One is that the model is one of the basic
``prototypic'' many-body models in condensed matter physics. It has become a 
testing ground for various analytical and numerical methods for strongly 
correlated systems, thus making it important to accurately establish its 
properties. Another reason is the very detailed predictions that have resulted
from field theoretical studies, such as renormalization group calculations 
for the nonlinear $\sigma$ model,\cite{csh,critical,neuberger,fisher} and 
chiral perturbation theory.\cite{hasenfratz} Apart from giving the low-energy 
properties in the thermodynamic limit, these theories also predict the system 
size dependence of various quantities.\cite{neuberger,fisher,hasenfratz} This
is important from the standpoint of numerical calculations such as exact
diagonalization and QMC, which are necessarily restricted to relatively small 
lattices. Finite-size scaling approaches have been very successful in the 
study of 1D quantum spin systems, having convincingly confirmed various
predictions from bosonization and conformal field theory. For example, 
critical exponents and logarithmic corrections have been extracted 
from the size dependence of ground state energies and finite-size 
gaps,\cite{lancscaling} and from correlation functions.\cite{dmrgscaling} 
With the concrete 
predictions now available, similar studies show great promise for testing 
theories also in 2D. For the standard Heisenberg model, finite-size scaling 
has been used extensively and successfully in extrapolating, e.g., the 
sublattice magnetization for small lattices to infinite system size,
\cite{reger,barnes,carlson,gross,trivedi,liang} but only a few studies have 
so far been accurate enough for reliably addressing the validity of the 
theoretical predictions for the size {\it corrections}.
\cite{runge1,runge2,wiese,beard} 

In one dimension, exact diagonalization, and more recently the density matrix 
renormalization group method,\cite{dmrg} enable highly accurate calculations 
for systems sufficiently large to approach the limit where the asymptotic 
scaling forms are valid.\cite{lancscaling,dmrgscaling} Calculations with these
methods in two dimensions cannot reach linear dimensions large enough to 
verify the details of the predicted scaling forms, however. Some of the 
expected leading finite-size behavior has been seen in exact diagonalization 
studies including systems with up to $6 \times 6$ spins,\cite{schulz1,schulz2} 
but constants extracted from the size dependence are typically not 
consistent with other calculations. For example, $c$ extracted from the 
scaling of $E$ deviates by 15\% from other estimates.\cite{schulz1} There 
are hence clear indications that these small systems are not yet in the 
regime where only the dominant corrections are important. 

QMC can reach significantly larger lattices at the cost of statistical errors 
which are often relatively large, making it difficult to accurately extract 
the scaling behavior. Runge carried out Green's function Monte Carlo (GFMC) 
simulations of $L \times L$ Heisenberg systems with $L$ up to $16$, and found 
a reasonable consistency with the leading $T=0$ size dependence of the energy 
and the sublattice magnetization.\cite{runge1,runge2} He also noted the 
presence of a subleading correction to the energy,\cite{runge2} but the 
accuracy of the GFMC data was not high enough to extract its order, and 
furthermore $c$ extracted from the leading correction was sensitive to the 
subleading one. The extrapolated ground state energy obtained in this study 
is nevertheless the most accurate estimate reported so far.\cite{runge2}

Chiral perturbation theory has recently enabled calculations of scaling forms
for finite size {\it and} finite temperature for various quantities.
\cite{hasenfratz} Such
forms have been used in combination with QMC data in recent work by Wiese 
and Ying,\cite{wiese} and Beard and Wiese.\cite{beard} Their calculations 
employed, respectively, methods based on the ``loop-cluster algorithm'' 
suggested by Evertz {\it et al.} \cite{evertz}, and a continuous-time variant 
of that method developed by Beard and Wiese.\cite{beard} These algorithms are 
based on global flips of loops of spins, and overcome the problems with long 
autocorrelation times typical of standard Suzuki-Trotter 
\cite{suzuki1,suzuki2} or worldline \cite{worldline} QMC methods (the 
continuous-time approach furthermore 
avoids the systematic discretization error of the Trotter approximation). 
Considerably more accurate finite-$T$ data could therefore be generated, and 
the leading-order scaling forms of chiral perturbation theory were 
convincingly verified, both in the ``cubic'' regime $T/c \approx 1/L$ 
(Ref.~\onlinecite{wiese}) and the ``cylindrical'' regime $T/c \ll 1/L$ 
(Ref.~\onlinecite{beard}). The extrapolated $M$, $\rho_s$ and $c$ are the 
most accurate reported to date, although there are some minor discrepancies 
between the two results for $M$ (on the border line of what could be 
expected within statistical errors alone).

In this paper, a finite-size scaling study of $T=0$ data is 
reported. Using the the Stochastic Series Expansion (SSE) QMC algorithm,
\cite{sse1,sse2} energy results of unprecedented accuracy are obtained for 
$L \times L$ lattices with $L$ up to $16$. The relative statistical errors 
are as low as $\approx 10^{-5}$. Employing a recently suggested data analysis 
scheme which takes into account covariance among calculated quantities,
\cite{covar} very accurate results for the sublattice magnetization are 
also obtained. Furthermore, the spin stiffness and the long-wavelength 
susceptibility $\chi (q=2\pi/L)$ are also calculated directly in the
simulations. 

Assuming for the size dependences polynomials in $1/L$, constrained by 
scaling forms for $E$ and $M$ predicted from chiral perturbation theory,
\cite{hasenfratz} all the computed quantities are included in a coupled 
$\chi^2$ fit. The quality of the QMC data for $E$ and $M$ is high enough 
that size corrections {\it beyond the subleading terms} have to be included. 
The leading-order corrections are fully consistent with the predictions.
From a careful statistical analysis of the fits, bounds for the subleading 
terms are estimated. The subleading energy correction is found to agree 
with the prediction of chiral perturbation theory to within a statistical 
error of 5\% (the subleading correction for $M$ is also estimated, but has
not yet been calculated analytically). This is the first numerical 
confirmation of chiral perturbation theory to subleading order.

The extrapolated ground state energy, $E=-0.669437(5)$, is the most accurate 
estimate reported to date, with a statistical error six times smaller than 
the GFMC result by Runge,\cite{runge2} and is slightly lower than his result.
Comparing the finite-size data of the two calculations, a clear tendency to 
over-estimation of the energy is seen in the GFMC results. This is likely 
due to a bias originating from ``population control'' in GFMC (a small effect 
of this nature was in fact anticipated by Runge\cite{runge2}). 

The results for the sublattice magnetization, $M=0.3070(3)$, and the 
spin-stiffness, $\rho_s=0.175(2)$, are both slightly lower than the estimates 
from the finite-$T$ scaling by Beard and Wiese\cite{beard} [$M=0.3083(2)$ and 
$\rho_s=0.185(2)$]. Although it is at this point difficult to definitely 
conclude which calculation is more reliable, it can again be noted that the 
high accuracy of the QMC data for $E$ and $M$ used in the fits carried out in 
this paper necessitates the inclusion of size corrections beyond the orders 
considered by Beard and Wiese.\cite{beard} Hence, any remaining effects of 
neglected corrections of even higher order should be smaller. Quantitative
estimates of such effects on the extrapolations performed here support that 
they are smaller than the statistical errors indicated above. For the spinwave
velocity the two calculations agree, the result obtained here being 
$c=1.673(7)$ and the value reported in Ref.~\onlinecite{beard} being 
$c=1.68(1)$.

The outline of the rest of the paper is the following. In Sec.~II the
SSE algorithm and the covariance error reduction scheme are outlined.
The absence of systematic errors are demonstrated in comparisons with exact 
results for $4\times 4$ and $6\times 6$ lattices. The fitting procedures and 
the results of these are discussed in Sec.~III. The study is summarized
in Sec.~IV. Some other problems where the methods applied here should be
useful are also mentioned.

\section{Numerical Methods}

The standard 2D Heisenberg model is defined by the Hamiltonian
\begin{equation}
\hat H = J\sum\limits_{\langle i,j\rangle} {\bf S}_i \cdot {\bf S}_j,
\quad (J > 0),
\end{equation}
where ${\bf S}_i$ is a spin-$1/2$ operator at site $i$ on a square lattice 
with $N=L\times L$ sites, and ${\langle i,j\rangle}$ denotes a pair of 
nearest-neighbor sites. Below, in II-A, the SSE approach to QMC simulation 
of this model is outlined. More details of the algorithm are discussed in 
Refs.~\onlinecite{sse2} and \onlinecite{chapter}. The SSE method 
has recently been applied to a variety of spin models,
\cite{bilayer1,bilayer2,dimrand,chains} as well as 1D Hubbard-type 
electronic models.\cite{ssefermions}

It was recently noted that correlations between measurements of different
observables can be used to significantly increase the accuracy of certain
quantities calculated in SSE simulations.\cite{covar} This covariance scheme 
for analyzing the data is of crucial importance in the present work, and 
therefore this method is also described below in Sec.~II-B. The high accuracy
of the procedures is demonstrated by comparing results for $4\times 4$ and
$6 \times 6$ lattices with the exact diagonalization data available for these
systems.

\subsection{Stochastic Series Expansion}

Based on the exact power series expansion of e$^{-\beta\hat H}$, the SSE 
method \cite{sse1,sse2} can be considered a generalization of Handscomb's 
QMC scheme.\cite{handscomb,lee} It is the first ``exact'' method proposed 
for QMC simulations of general lattice Hamiltonians at finite-temperature
(with the usual caveat of being in practice restricted to models for which
the sign problem can be avoided). It is hence not based on a controlled 
approximation, such as the Trotter formula used in standard worldline methods,
\cite{worldline} and directly gives results accurate to within statistical 
errors. Despite being formulated at finite $T$, temperatures low enough for 
studying the ground state can easily be reached for moderate-size lattices.

As in Handscomb's method for the $S=1/2$ antiferromagnet,\cite{lee}
the SSE approach for this model starts from the Hamiltonian written as 
($J=1$)
\begin{equation}
\hat H = -{1\over 2} \sum\limits_{b=1}^{2N} [\hat H_{1,b} - \hat H_{2,b}]
+ {N\over 2},
\label{ham2}
\end{equation}
where $b$ is a bond connecting a pair of nearest-neighbor sites
$\langle i(b),j(b)\rangle$, and the operators $\hat H_{1,b}$
and $\hat H_{2,b}$ are defined as
\begin{mathletters}
\begin{eqnarray}
\hat H_{1,b} & = & 2[\hbox{$1\over 4$} - S^z_{i(b)}S^z_{j(b)}] \\
\hat H_{2,b} & = & S^+_{i(b)}S^-_{j(b)} + S^-_{i(b)}S^+_{j(b)} .
\end{eqnarray}
\end{mathletters}
An exact expression for an operator expectation value 
\begin{equation}
\langle \hat A \rangle = {1\over Z}
{\rm Tr}\lbrace \hat A {\rm e}^{-\beta \hat H} \rbrace ,\quad
Z = {\rm Tr}\lbrace {\rm e}^{-\beta \hat H} \rbrace ,
\end{equation}
at inverse temperature $\beta = J/T$, is obtained by Taylor expanding
e$^{-\beta \hat H}$ and writing the traces as  sums over diagonal matrix
elements in the basis $\lbrace |\alpha \rangle =
|S^z_1,\ldots,S^z_N \rangle \rbrace$. The partition function is then
\cite{sse1}
\begin{equation}
Z = \sum\limits_\alpha \sum\limits_{n=0}^\infty \sum\limits_{S_n}
{(-1)^{n_2} \beta^n \over n!} \left \langle \alpha \left | 
\prod\limits_{i=1}^n \hat H_{a_i,b_i} \right | \alpha \right \rangle ,
\label{partition}
\end{equation}
where $S_n$ is a sequence of index pairs
defining the operator string $\prod_{i=1}^n \hat H_{a_i,b_i}$,
\begin{equation}
S_n = [a_1,b_1][a_2,b_2]\ldots [a_n,b_n],\quad
a_i \in \lbrace 1,2\rbrace$, $b_i \in \lbrace 1,\ldots ,2N \rbrace,
\label{sn}
\end{equation}
and $n_2$ denotes the total number of index pairs (operators) $[a_i,b_i]$ 
with $a_i = 2$ ($n=n_1+n_2$). Eq.~(\ref{partition}) deviates from Handscomb's
method,\cite{handscomb,lee} which relies on exact evaluation of the traces 
of the operator sequences and therefore is limited to models for which this 
is possible. The Heisenberg model considered here is such a model (one of
the very few), but the more general SSE approach of further expanding over 
a set of basis states is preferable also in this case, for reasons that will 
be discussed below.

The objective now is to develop a scheme for importance sampling of the terms 
in the partition function (\ref{partition}). A term, or configuration,
$(\alpha ,S_n)$ is specified by a basis state $| \alpha \rangle$ and an 
operator sequence $S_n$.
The operators $\hat H_{1,b}$ and $\hat H_{2,b}$ can act only on states where 
the spins at sites $i(b)$ and $j(b)$ are antiparallel. The diagonal 
$\hat H_{1,b}$ leaves such a state unchanged, whereas the off-diagonal 
$\hat H_{2,b}$ flips the spin pair. Defining a propagated state
\begin{equation}
| \alpha (p) \rangle = \prod\limits_{i=1}^p \hat H_{a_i,b_i} |\alpha \rangle ,
\quad | \alpha (0) \rangle = | \alpha \rangle ,
\label{propagated}
\end{equation}
a configuration $(\alpha ,S_n)$ must clearly satisfy the periodicity
condition  $|\alpha (n) \rangle = | \alpha (0) \rangle$ in order to
contribute to the partition function. For a lattice with $L \times L$ sites
and $L$ even, this implies that the total number $n_2$ of the off-diagonal 
operators must be even, and hence that all terms in (\ref{partition}) are 
positive and can be used as relative probabilities in a Monte Carlo importance
sampling procedure (this is true for any non-frustrated system).

For a finite system at finite $\beta$, the powers $n$ contributing 
significantly to the partition function are restricted to
within a well defined regime, and the sampling space is therefore finite in 
practice. In order to construct an efficient updating scheme for the 
index sequence it is useful to explicitly truncate the Taylor expansion
at some self-consistently chosen upper bound $n=l$, high enough to 
cause only an exponentially small, completely negligible error.\cite{sse1} 
One can then define a sampling space where the length of the sequence is
{\it fixed}, by inserting a number $l-n$ of unit operators, denoted 
$\hat H_{0,0}$, in the operator strings. The terms in the partition 
function (\ref{partition}) are divided by ${l \choose n}$, in order to 
compensate for the number of  different ways of inserting the unit 
operators. The summation over $n$ in
(\ref{partition}) is then implicitly included in the summation over
all sequences $S_l$ of length $l$, with $[a_i ,b_i]=[0,0]$ as
an allowed operator. Denoting by $W(\alpha ,S_l)$ the
weight of a configuration $(\alpha ,S_l)$, the partition
function is then
\begin{equation}
Z = \sum\limits_{\alpha} \sum\limits_{S_l} W(\alpha ,S_l) .
\end{equation}
Since all non-zero matrix elements in (\ref{partition})
equal one, the weight is (when non-vanishing)
\begin{equation}
W(\alpha ,S_l ) = \left ({\beta\over 2} \right )^n {(l-n)!\over l!},
\label{wl}
\end{equation}
where $n$ still is the expansion power of the term, i.e., the  number of 
non-$[0,0]$ operators in $S_l$.

The following is only a brief outline of the actual sampling scheme. More
details can be found in Refs.~\onlinecite{sse2} and \onlinecite{chapter}.
During the simulation, $S_l$ and one of the states $|\alpha (p)\rangle$
are stored. Other propagated states are generated as needed. The simulation 
is started with a randomly generated state $| \alpha (0)\rangle$, with an 
index sequence $S_l$ containing only $[0,0]$ operators (unit operators), and 
with some arbitrary (small) $l$. The truncation $l$ is adjusted as the 
simulation proceeds, as will be discussed further below. 

With the fixed-length scheme, all updates of the operator sequence can be 
formulated in terms of substitutions of one or several operators. The simplest
involves a diagonal operator at a single position; $[0,0] \leftrightarrow 
[1,b]$. This update can be carried out consecutively at all positions $p$ 
for which $a_p \in \lbrace 0,1\rbrace$. In the $\rightarrow$ direction, the 
bond index $b$ is chosen at random and the update is rejected if the spins 
connected by $b$ are parallel in the current state $|\alpha (p-1)\rangle$. 
The Metropolis acceptance probabilities \cite{metropolis} required to satisfy 
detailed balance are obtained from (\ref{wl}), where the power $n$ in is 
changed by $\pm 1$. Updates involving the off-diagonal operators $[2,b]$ are 
carried out with $n$ fixed. The simplest is of the type $[1,b][1,b] 
\leftrightarrow [2,b][2,b]$, involving two operators acting on 
the same bond. These two sequence updates can generate all configurations
with spin flips on retracing paths on the lattice, and are the only ones 
required for a 1D system with open boundary conditions. For a 2D system, 
configurations associated with spin flips around any closed loop are possible,
and an additional type of update is required. It is sufficient to consider 
substitutions on a plaquette, of the type $[2,b_1][2,b_2] 
\leftrightarrow [2,b_3][2,b_4]$, where $b_1 ,\ldots, b_4$ is a permutation 
of the four bonds of a plaquette. For systems with periodic boundary 
conditions, updates involving cyclic spin flips on loops wrapping around the 
whole system are required (sampling of different winding number sectors), 
and cannot be accomplished by the above local sequence alterations. 
For the square lattice considered here, the winding number can be changed
by substituting $L/2$ operators according to $[2,b_1]\ldots [2,b_{L/2}] 
\leftrightarrow [2,b_{L/2+1}]\ldots [2,b_L]$, where the set of bonds 
$b_1 ,\ldots ,b_L$ is a permutation of bonds forming a closed ring around 
the system in the $x$- or $y$-direction. 

Updating the operator sequence with the four types of operator substitutions 
described above suffices for generating all possible configurations within 
a sector of fixed magnetization, $m^z=\sum_{i=1}^N S^z_i$. In the grand 
canonical ensemble, global spin flips changing the magnetization are also
required. Here $T \to 0$ will be considered (i.e., $T$ is much lower than 
the finite-size gap), and since the ground state is a singlet\cite{lieb} 
the canonical ensemble with $m^z=0$ is appropriate. It can be noted that 
in Handscomb's method the sampling is (in principle) automatically over all 
magnetization sectors, and therefore a restriction to, e.g., $m^z=0$ is not 
possible. In practice, this causes problems at low temperatures, and 
Handscomb's method has therefore been used for the antiferromagnetic 
Heisenberg model mostly at relatively high temperatures ($T/J \agt 0.4$ 
in 2D).\cite{lee} The SSE method with the restriction $m^z=0$ can be used 
at arbitrarily low $T$.

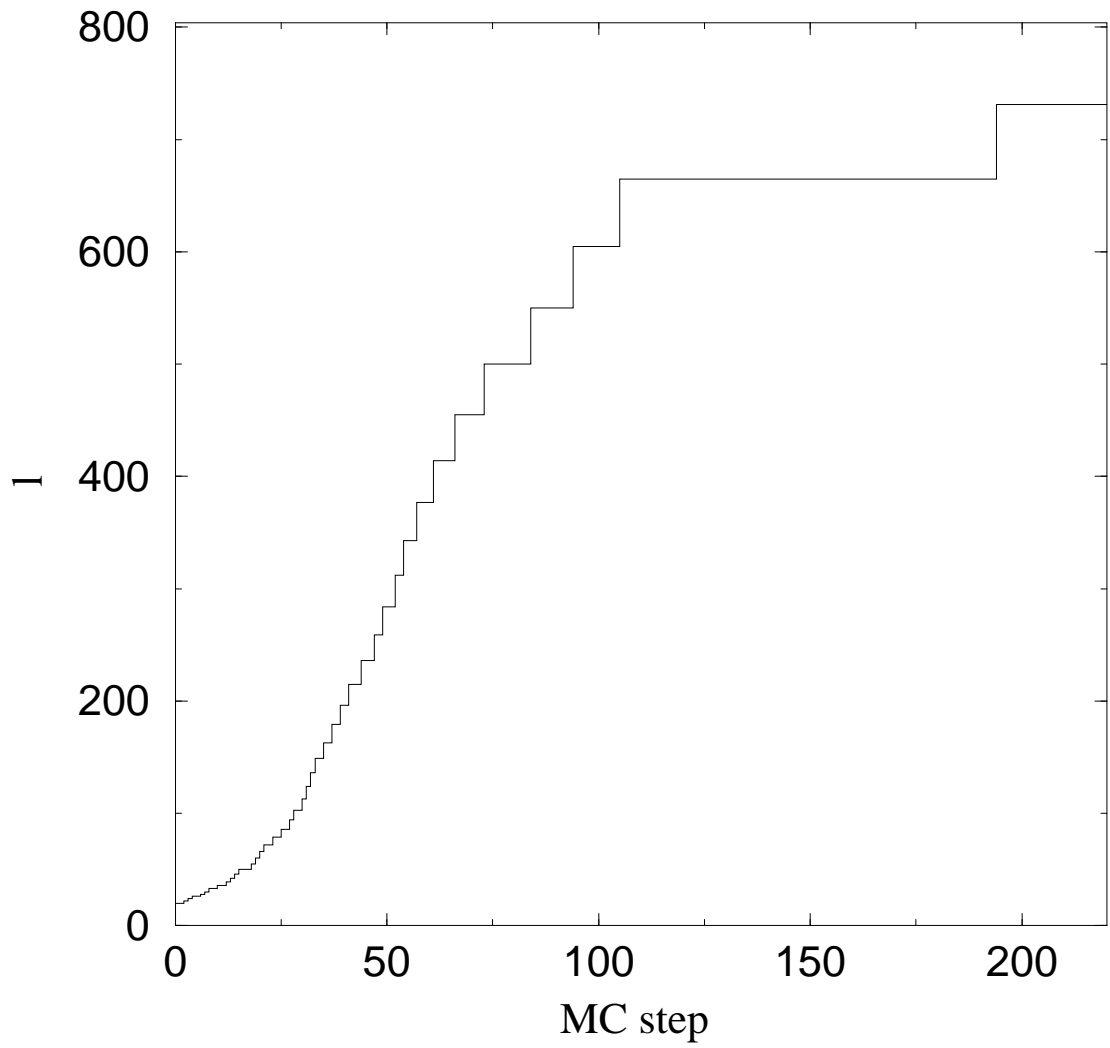
\begin{figure}
\caption{The operator sequence truncation vs.~the number of Monte Carlo steps 
performed for a $4\times 4$ lattice at inverse temperature $\beta=32$. The 
final $l$ after $10^5$ MC steps was $l=804$ (the last increase
occurred after 6674 steps). The increment used was $\Delta_l=l/10$.}
\label{figlen}
\end{figure}

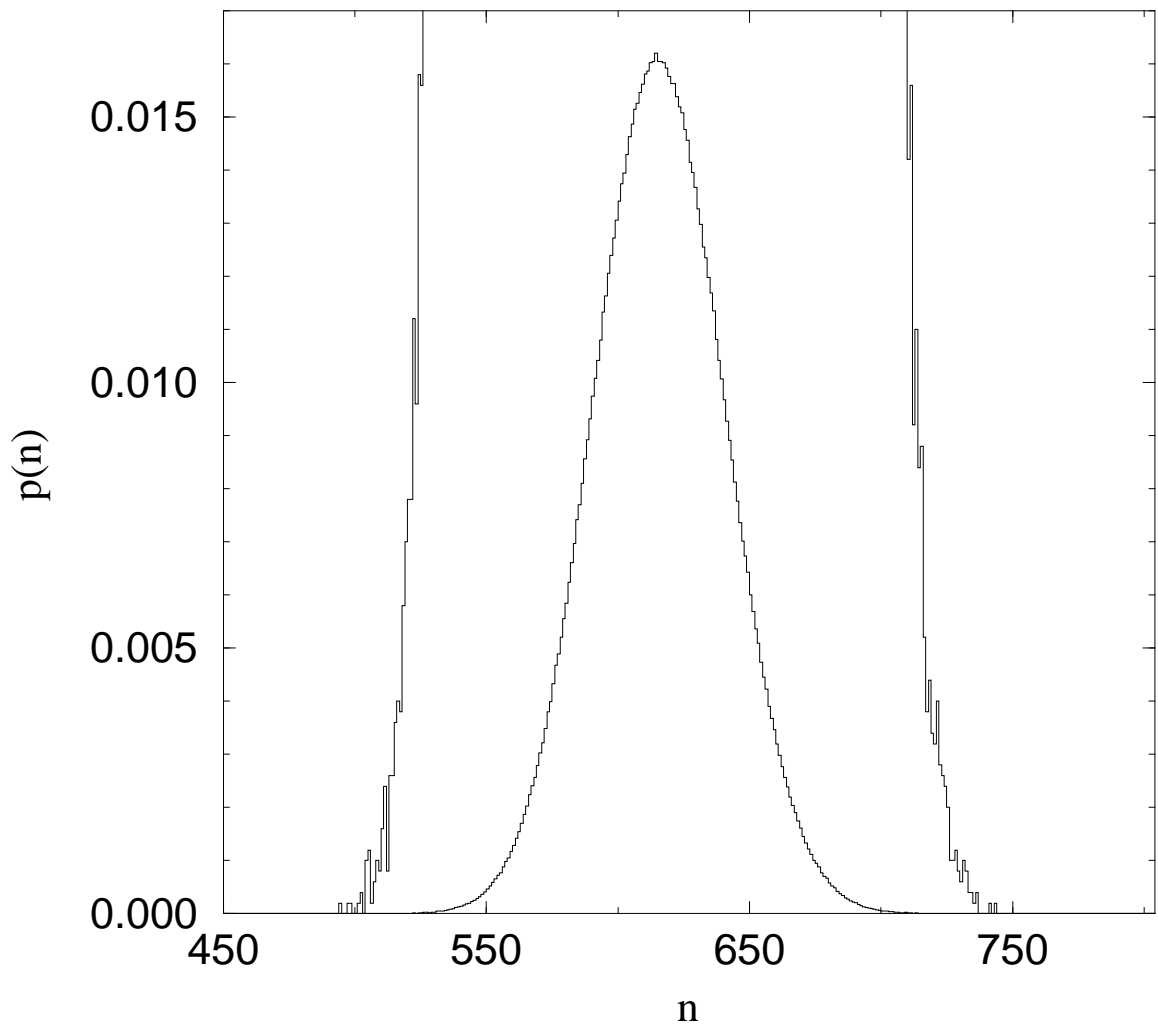
\begin{figure}
\caption{The distribution of the power $n$ of the sampled terms in a 
$5 \times 10^6$ MC step simulation of $4\times 4$ lattice at inverse 
temperature $\beta=32$, after adjusting $l$ as shown in Fig.~1. The lower 
histogram is the full distribution. The higher, only partially visible
histogram is the distribution multiplied by a factor $1000$. The cut-off was
$l=804$, which is significantly larger than the largest $n$ sampled. Hence,
the truncation has not degraded the accuracy of the simulation.}
\label{figdist}
\end{figure}

In order to determine a sufficiently high truncation of the expansion, the 
fluctuating power $n$ is monitored during the equilibration part of the 
simulation. If $n$ exceeds some threshold $l-\Delta_l/2$, the cut-off is 
increased, $l \to l + \Delta_l$, by inserting additional $\hat H_{0,0}$ 
operators at random positions. In practice $\Delta_l \approx l/10$ leads
to a rapid saturation of $l$ at a value sufficient to cause no 
detectable truncation errors. The growth of $l$ during equilibration is 
illustrated for a $4 \times 4$ system in Fig.~\ref{figlen}. The distribution 
of $n$ during a subsequent simulation is shown in Fig.~\ref{figdist}, and 
clearly demonstrates that the truncation of the expansion is no approximation
in practice.

A Monte Carlo step (MC step) is defined as a series of the 
single-(diagonal)operator substitutions attempted consecutively at each 
position in $S_l$ (where possible), followed by a series of off-diagonal
updates carried out on each bond, plaquette, and ring. Due to the locality 
of the constraints in these updates, the number of operations (the CPU time) 
per MC step scales linearly with $N$ and $\beta$.\cite{sse2,chapter} 
However, the acceptance rate for the
``ring update'' that changes the winding number decreases rapidly with 
increasing system size. It is therefore sometimes useful to increase the 
number of attempted ring updates with the system size, which then leads to a 
faster growth of CPU time with $N$. The acceptance rate of the ring update 
currently used becomes too low for $L \agt 16$, and simulations of larger 
systems therefore in practice have to be restricted to the sector with 
zero winding number. It has recently been noted\cite{patrik} that in fact 
exact results are obtained as $T \to 0$ even for simulations restricted this 
way. However, compared to simulations with fluctuating winding numbers, lower 
temperatures are required for the system observables to saturate at their 
ground state values.\cite{patrik} Here only systems with $L \le 16$ are 
considered, and the update changing the winding number is always included.

Measurements of physical observables are carried out using the index 
sequences $S_n$ obtained by omitting the $[0,0]$ operators in the generated
$S_l$. These are then, of course, distributed according to the weight function 
corresponding to Eq.~(\ref{partition}). 

One can show that the internal energy per spin is simply given by the average 
of $n$ [with the constant term in Eq.~(\ref{ham2}) neglected]:
\cite{handscomb,sse1}
\begin{equation}
E = -{\langle n\rangle \over N \beta} .
\label{energy}
\end{equation}
This expression also shows that the average power, and hence the sequence
length $l$, scales as $\beta N$ at low temperatures. 

A spin-spin correlation function,
\begin{equation}
C(i,j)=C({\bf r}_i-{\bf r}_j)=\langle S^z_i S^z_j \rangle ,
\end{equation}
is obtained averaging the correlations in the propagated states
$| \alpha (p) \rangle$ defined in Eq.~(\ref{propagated}). Further defining
\begin{equation}
S^z_i[p]= \langle \alpha (p) |  S^z_i | \alpha (p) \rangle,
\end{equation}
the correlation function is given by \cite{sse1}
\begin{equation}
C(i,j) = \left \langle 
{1\over n+1} \sum\limits_{p=0}^n S^z_i[p]S^z_j[p]
\right \rangle .
\label{correl}
\end{equation}
The corresponding static susceptibility,
\begin{equation}
\chi (i,j) = \int\limits _0^\beta d\tau 
\langle S^z_i (\tau) S^z_j (0) \rangle ,
\end{equation}
involves correlations between all the propagated states:\cite{sse2}
\begin{equation}
\chi (i,j)=\left \langle {\beta\over n(n+1)} 
\left ( \sum\limits_{p=0}^{n-1} S^z_i[p] \right )
\left ( \sum\limits_{p=0}^{n-1} S^z_j[p] \right )
+ {\beta\over (n+1)^2} 
\left ( \sum\limits_{p=0}^{n} S^z_i[p]S^z_j[p] \right )
\right \rangle .
\label{susc}
\end{equation}

Off-diagonal correlation functions can be easily calculated for operators
that can be expressed in terms of the spin-flipping operators $\hat H_{2,b}$, 
each of which is a sum of two terms; $\hat H^+_b = S^+_{i(b)}S^-_{j(b)}$ and 
$\hat H^-_b=S^-_{i(b)}S^+_{j(b)}$. The spin stiffness constant involves a 
static susceptibility defined in terms of these operators.

Although the simulation scheme is formulated with $\hat H_{2,b}=H^+_b + 
H^-_b $, one can still access the terms individually since only one of them 
can propagate a given state. One can show that an equal-time correlation 
function, 
\begin{equation}
F_{\sigma \sigma'} (b,b') = 
\langle \hat H^{\sigma}_{b} \hat H^{\sigma '}_{b'} \rangle ,
\end{equation}
is given by \cite{sse2}
\begin{equation}
F_{\sigma \sigma'} (b,b') = \left \langle {n-1\over (\beta/2)^2 } 
N(b\sigma ; b'\sigma ') \right \rangle ,
\label{equaloff}
\end{equation}
where $N(b\sigma ; b'\sigma ')$ is the number of times the operators
$\hat H^{\sigma}_{b}$ and  $\hat H^{\sigma '}_{b'}$ appear next to each
other in $S_n$, in the given order. The corresponding static susceptibility,
\begin{equation}
\chi_{\sigma \sigma'} (b,b') = \int\limits_0^\beta d\tau
\langle \hat H^{\sigma}_{b}(\tau) \hat H^{\sigma '}_{b'}(0) \rangle ,
\label{offsusdef}
\end{equation}
is given by the remarkably simple formula\cite{sse2}
\begin{equation}
\chi_{\sigma \sigma'} (b,b') = 4
\left \langle N(b\sigma)  N(b'\sigma') 
- \delta_{bb'} \delta_{\sigma\sigma'} N(b\sigma)  \right \rangle / \beta,
\label{offsus}
\end{equation}
where $N(b\sigma)$ is the total number of operators $\hat H_b^\sigma$ in 
$S_n$. 

Now a direct estimator for the spin stiffness can be constructed. The
stiffness, $\rho_s$, is defined as the second derivative of the ground state 
energy with respect to a twist $\Phi$ in the boundary condition, around
an axis perpendicular to the direction of the broken symmetry. For a finite
lattice, where the symmetry is not broken, a factor $3/2$ has to be included 
in order to account for rotational averaging. Distributing the twist equally 
over all interacting spin pairs $\langle i,j\rangle_x$ in the $x$-direction, 
the finite-size definition for $\rho_s$ is hence
\begin{equation}
\rho_s = {3\over 2}
{1\over L^2} {\partial^2 E_0 (\phi) \over \partial \phi^2} \Bigl | _{\phi=0},
\label{stiffderiv}
\end{equation}
where $\phi = \Phi/L$. An expression which is only dependent on
the ground state at $\phi=0$ is obtained by expanding the Hamiltonian
to second order in $\phi$. The Hamiltonian in the presence of the twist is
\begin{equation}
\hat H(\phi) = 
\sum\limits_{~\langle i,j\rangle_x} {\bf S}_i \cdot R(\phi) {\bf S}_j +
\sum\limits_{~\langle i,j\rangle_y} {\bf S}_i \cdot {\bf S}_j,
\end{equation}
where $R(\phi)$ is the rotation matrix
\begin{equation}
R(\phi) =  \left ( \begin{array}{ccc}
 \cos{(\phi)} & \sin{(\phi)} & 0 \\
-\sin{(\phi)} & \cos{(\phi)} & 0 \\
    0         &      0       & 1
\end{array} \right ).
\end{equation}
Expanding to second order in $\phi$ results in
\begin{equation}
\hat H (\phi) - \hat H (0) =
-\frac{1}{2} \sum\limits_{~\langle i,j\rangle_x} \left [
\phi^2 (S^x_iS^x_j + S^y_iS^y_j) + 
i\phi (S^+_iS^-_j - S^-_iS^+_j) \right ].
\end{equation}
The first term is proportional to $\hat H(0)$ (for the rotationally invariant
case considered here). The expectation value of the second term vanishes, but 
it gives a contribution quadratic in $\phi$ in second order perturbation 
theory. Defining the spin current operator
\begin{equation}
j_s = \frac{i}{2}\sum\limits_{~\langle i,j\rangle_x} 
(S^+_iS^-_j - S^-_iS^+_j),
\end{equation}
and the current-current correlation function at Matsubara frequency
$\omega_m = 2\pi mT$,
\begin{equation}
\Lambda _s (\omega_m)= \frac{1}{L^2} \int\limits_0^\beta d\tau 
{\rm e}^{-i\omega_m \tau} \langle j_s (\tau) j_s (0) \rangle ,
\label{lambdas}
\end{equation}
the stiffness is given by
\begin{equation}
\rho_s = - \hbox{$3\over 2$} [\hbox{$1\over 3$}E + \Lambda _s (0)] ,
\label{stiffdef}
\end{equation}
where $E$ is the ground state energy per spin. 

The QMC estimate for the energy is given by Eq.~(\ref{energy}). The 
current-current correlator $\Lambda_s \equiv \Lambda_s (0)$ is a sum 
of integrals of the form (\ref{offsusdef}). Denoting by $N^+_x$  and $N^-_x$ 
the number in $S_n$ of operators $S^+_iS^-_j$ and $S^-_iS^+_j$ with 
$\langle i,j\rangle$ a bond in the $x$-direction, Eqs.~(\ref{lambdas}) and 
(\ref{offsus}) give
\begin{equation}
\rho_s = {3/2 \over \beta N }
\left \langle ( N^+_x - N^-_x)^2 \right\rangle  ,
\end{equation}
i.e. the terms linear in $N^+_x$ and $N^-_x$ cancel. Defining
the winding numbers $w_x$ and $w_y$ in the $x$ and $y$ direction:
\begin{equation}
w_\alpha = (N^+_\alpha - N^-_\alpha) \bigr / L ,\quad (\alpha = x,y),
\end{equation}
the stiffness can also be written as
\begin{equation}
\rho_s = \hbox{$3\over 4$}
\left \langle w_x^2 + w_y^2 \right \rangle \bigr / \beta.
\label{rhow}
\end{equation}
This definition is clearly valid only for a simulation that samples all 
winding number sectors. With a restriction to the subspace with $w_x = w_y = 
0$, $\rho_s$ can be calculated using the long-wavelength limit of a 
current-current correlator involving a twist field with a spatial 
modulation.\cite{scalapino} 

The above method of calculating the stiffness directly from the winding
number fluctuations is clearly strongly related to methods used for the 
superfluid density in simulations of boson models.
\cite{pollock}

\subsection{Error Reduction Using Covariance}

In Monte Carlo simulations, fluctuations (statistical errors) of different 
physical observables are often correlated with each other. These covariance 
effects can sometimes be used to obtain improved estimators for certain 
quantities.\cite{covar} In some cases one may have exact knowledge of some 
quantity independently of the QMC calculation. If there are strong 
correlations between a known quantity $A$ and some other, unknown quantity 
$B$, the accuracy of $B$ can be improved via its covariance with the measured 
$A$, by calculating the average and statistical error under the condition that
$A$ equals its known value. In other cases, it may be possible to calculate a 
quantity in more than one way in the same simulation. If one of the estimates,
$A_1$, is more accurate than the other, $A_2$, a covariance between $A_2$ and 
some other quantity $B$ can again be used to improve the estimate of $B$. 
With the SSE method the internal energy of the rotationally invariant 
Heisenberg model can be calculated in two different ways: $E_1$ from the 
average power of the series expansion according to Eq.~(\ref{energy}), 
and using the nearest-neighbor correlation function $C(1,0)$ calculated 
according to Eq.~(\ref{correl}); $E_2=6C(1,0)$. The manifestly rotationally 
invariant estimator $E_1$ is significantly less noisy than $E_2$. Results 
for quantities with fluctuations correlated to those of $C(1,0)$, such as 
$C({\bf r})$ with $r > 1$, can therefore be improved with the aid of $E_1$. 

For the purpose of accurately measuring correlations between the fluctuations 
of two different quantities, the so called ``bootstrap method'' is a useful 
tool.\cite{bootstrap} With the simulation data as usually divided into $M$ 
``bin'' averages, a  ``bootstrap sample'' $\bar A_R$ is defined as an average 
over $M$ randomly selected bins (i.e., the same number as the total number 
of bins, allowing, of course, multiple selections of the same bin). With 
$r(i)$ denoting the $i$:th randomly chosen bin,
\begin{equation}
\bar A_R  = {1 \over M} \sum\limits_{i=1}^M A_{r(i)} .
\label{randomav}
\end{equation}
The statistical error can be calculated on the basis of $M_R$ bootstrap 
samples $\bar A_{R_i}$, according to\cite{bootstrap}
\begin{equation}
\sigma^2  = {1 \over M_R} \sum\limits_{i=1}^{M_R} 
(\bar A_{R_i} - \bar A )^2 ,
\label{booterror}
\end{equation}
where $\bar A$ is the regular average over all bins. Note that 
Eq.~(\ref{booterror}) lacks the factor $(M_R-1)^{-1}$ present in the 
conventional expression for the variance of the average calculated 
on the basis of $M_R$ bins. The bootstrap method is in general more accurate
(due to a better realization of a Gaussian distribution for the bootstrap
samples), in particular if $A$ is 
not measured directly in the simulation, but is some nonlinear function of 
measured quantities (in which case $A$ should be calculated on the basis of 
bootstrap samples, not individual bins). Sets of bootstrap samples $\lbrace 
\bar A_{R_i} \rbrace$ and $\lbrace \bar B_{R_i} \rbrace$ generated on the 
basis of the same randomly selected bins are well suited for evaluating 
correlations between the statistical fluctuations of $A$ and $B$, and are 
used in the covariance error reduction scheme described next.

Here this method will be illustrated using simulation results for the 
staggered structure factor $S(\pi,\pi)$ and the staggered susceptibility 
$\chi(\pi ,\pi)$. These are defined according to
\begin{mathletters}
\begin{eqnarray}
S(\pi,\pi) &=& {1\over N} \sum\limits_{i,j} 
(-1)^{x_j-x_i+y_j-y_i}C(i,j)
\label{spi} \\
\chi(\pi,\pi)&=&{1\over N} \sum\limits_{i,j} 
(-1)^{x_j-x_i+y_j-y_i}\chi (i,j),
\label{xpi}
\end{eqnarray}
\end{mathletters}
with $C(i,j)$ and $\chi (i,j)$ given by Eqs.~(\ref{correl}) and (\ref{susc}).
The structure factor is of particular interest, since it defines the 
sublattice magnetization squared of a finite system. The fluctuations of 
$S(\pi ,\pi)$ are strongly correlated to those of $C(1,0)$, and $S(\pi ,\pi)$
can therefore be calculated to an accuracy significantly higher than if only 
the direct estimator (\ref{correl}) is used. The susceptibility 
$\chi(\pi,\pi)$ is only weakly correlated with $C(1,0)$, however, and only 
a modest gain in accuracy can be achieved for this quantity.

First some results for a $6\times 6$ lattice are discussed. This is the largest
system for which Lanczos results have been obtained.\cite{schulz2} Comparing 
with these exact results, the accuracy of the QMC technique and the covariance
method can be rigorously checked. The temperature used in the simulation 
has to be low enough for the calculated quantities to have saturated at their 
ground state values. In order to check for temperature effects, several 
calculations were carried out. Results at inverse temperatures $\beta=24$ and 
$48$ are indistinguishable within error bars, indicating that these 
temperatures are sufficiently low for $L=6$. The results presented below are 
for $\beta=48$. The simulation was divided into bins of $5 \times 10^5$ MC 
step each, and a total of $600$ bins were generated.

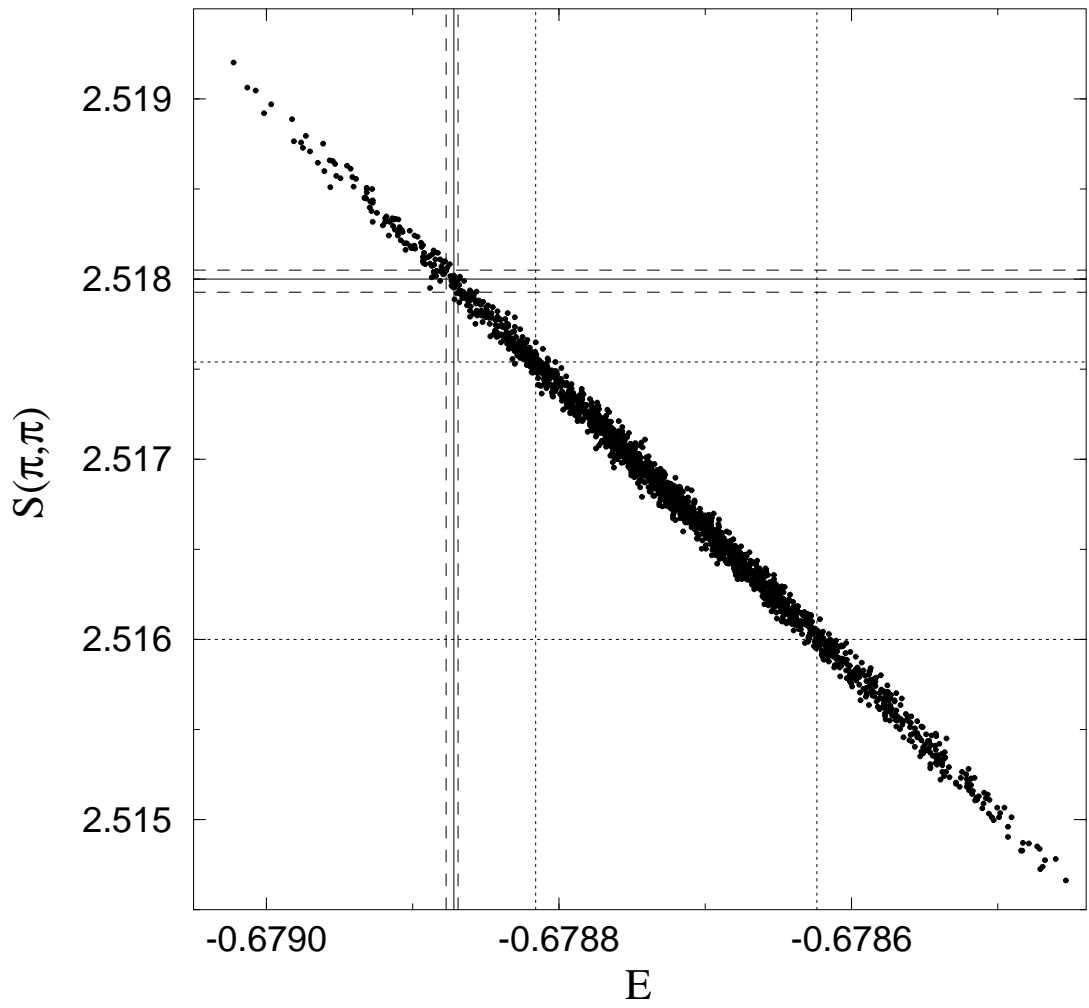
\begin{figure}
\caption{Correlations between the staggered structure factor and the energy as 
obtained from the nearest-neighbor correlation function for a $6\times 6$ 
lattice at $\beta=48$. Each point represents a bootstrap sample of QMC bin 
averages. The 
solid vertical line indicates the exact energy, and the solid horizontal line 
is the exact structure factor [the result for $S(\pi,\pi)$ is given ``only'' 
with 5 significant digits in Ref.~\protect{\onlinecite{schulz2}}, which 
actually implies a small uncertainty on the scale of this figure]. 
The vertical dotted and dashed lines indicate, respectively, the estimate 
$\pm$ one standard deviation of the energy calculated from the 
nearest-neighbor correlation function and the average of $n$. The dotted
horizontal lines indicate the conventional estimate of the structure factor,
and the dashed ones the improved result obtained using the covariance 
method.}
\label{figcov6}
\end{figure}

Fig.~\ref{figcov6} shows the covariance between the measured nearest-neighbor
correlation function ($E_2$) and $S(\pi,\pi)$. The plot was generated on  
the basis of $2000$ bootstrap samples. Strong linear correlations between 
the two quantities are evident. Hence further knowledge of $E$ can improve 
the estimate of $S$. The conventional average and error of 
$S(\pi,\pi)$ is calculated on the basis of all the points i.e., the 
distribution obtained by projecting the points onto the $S$-axis. Having a 
better estimate $E_1 \pm \sigma_1$ for $E$, an improved estimate of $S$ can 
be calculated by weighting the points by a Gaussian centered at $E_1$ and 
with a width equal to the error $\sigma_1$. In this case, the reduced 
statistical error is $\approx 1/12$ of the conventional error. Note that the 
conventional estimates of both $S$ and $E_2$ lie outside the exact results 
by $\approx 1.5$ standard deviations (not an unlikely situation statistically).
The improved estimate of $S$ is nevertheless within one standard deviation 
of the exact result, reflecting this being the case for the more accurate 
energy estimate $E_1$ used in the procedure. In fact, this correcting property
of the covariance method can even eliminate certain systematic errors, such 
as those originating from finite-$T$ effects in calculations aimed at ground 
state properties.
\cite{covar}

Besides illustrating the use of the covariance method to reduce the
statistical errors, Fig.~\ref{figcov6} also clearly demonstrates to a high
accuracy the absence of detectable systematic errors in the QMC data. 
This confirms that the SSE method indeed produces unbiased results. 
Table~\ref{tab1} summarizes the comparisons with the exact results for 
both $4\times 4$ and $6\times 6$ lattices.

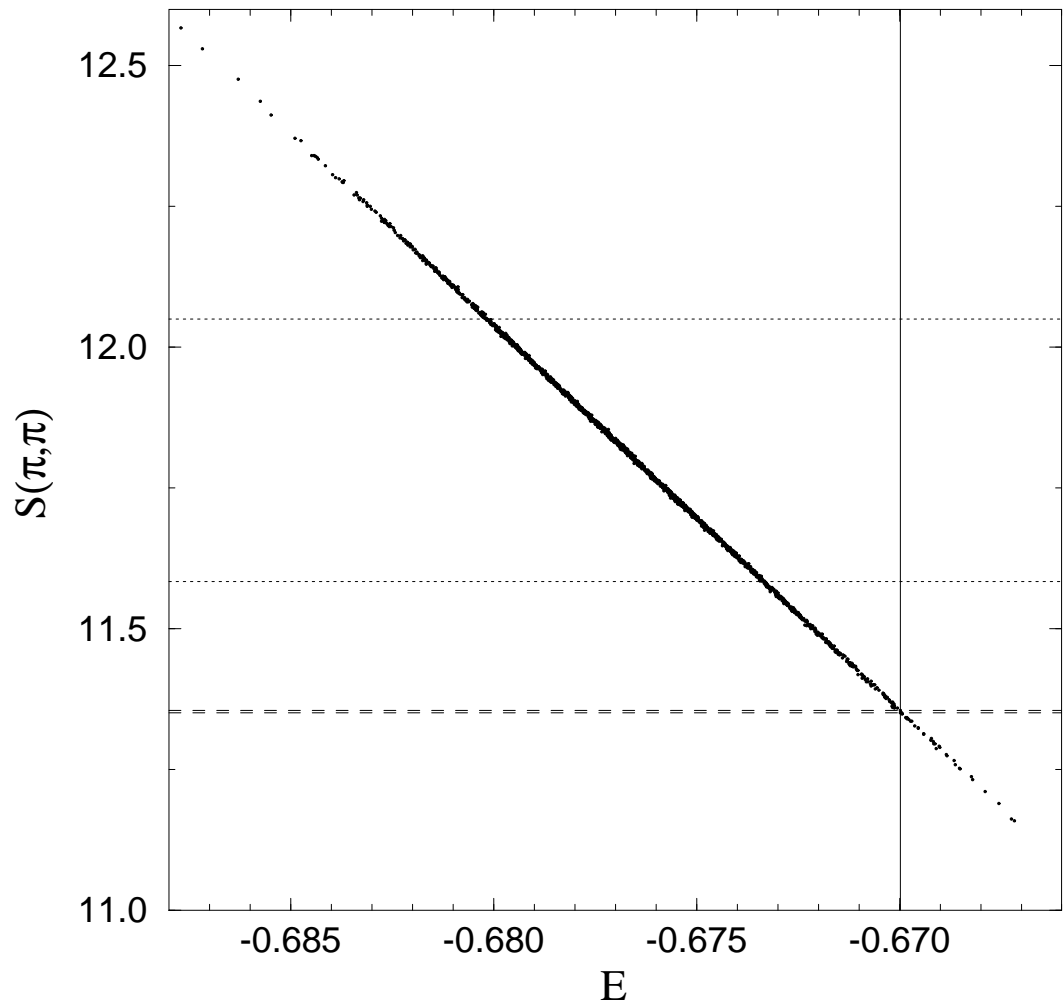
\begin{figure}
\caption{Correlations between the energy as obtained from the nearest-neighbor
correlation function and the staggered structure factor for a $16\times 16$ 
lattice at $\beta=128$. The vertical line indicates the estimate of the energy
from the average of $n$ (the error is too small to be seen on this scale). 
The dotted horizontal lines indicate the value $\pm$ one standard deviation of 
$S(\pi,\pi)$ calculated using the conventional estimator. The dashed,
almost indistinguishable lines indicate the improved estimate.}
\label{figcov16}
\end{figure}

As the system size increases, the fluctuations in $S(\pi,\pi)$ as computed in 
the standard way increase, and accurate estimates become increasingly 
difficult to obtain. This is typical of algorithms utilizing local updates.
The fluctuations in the energy per site as calculated from $\langle n\rangle$ 
actually decrease, however (due to self-averaging). Hence, the gain in 
accuracy achieved with the covariance effect increases with the system 
size. Fig.~\ref{figcov16} shows $L=16$ results for the staggered structure 
factor. For this system size the error in the energy estimate $E_1$ is 
negligible on the scale of the fluctuations of $E_2$, and the error in 
the improved $S(\pi,\pi)$ is essentially the width of the elongated shape in 
the vertical direction. In this case the covariance method leads to error 
bars $\approx 1/100$ of those calculated in the standard way. 

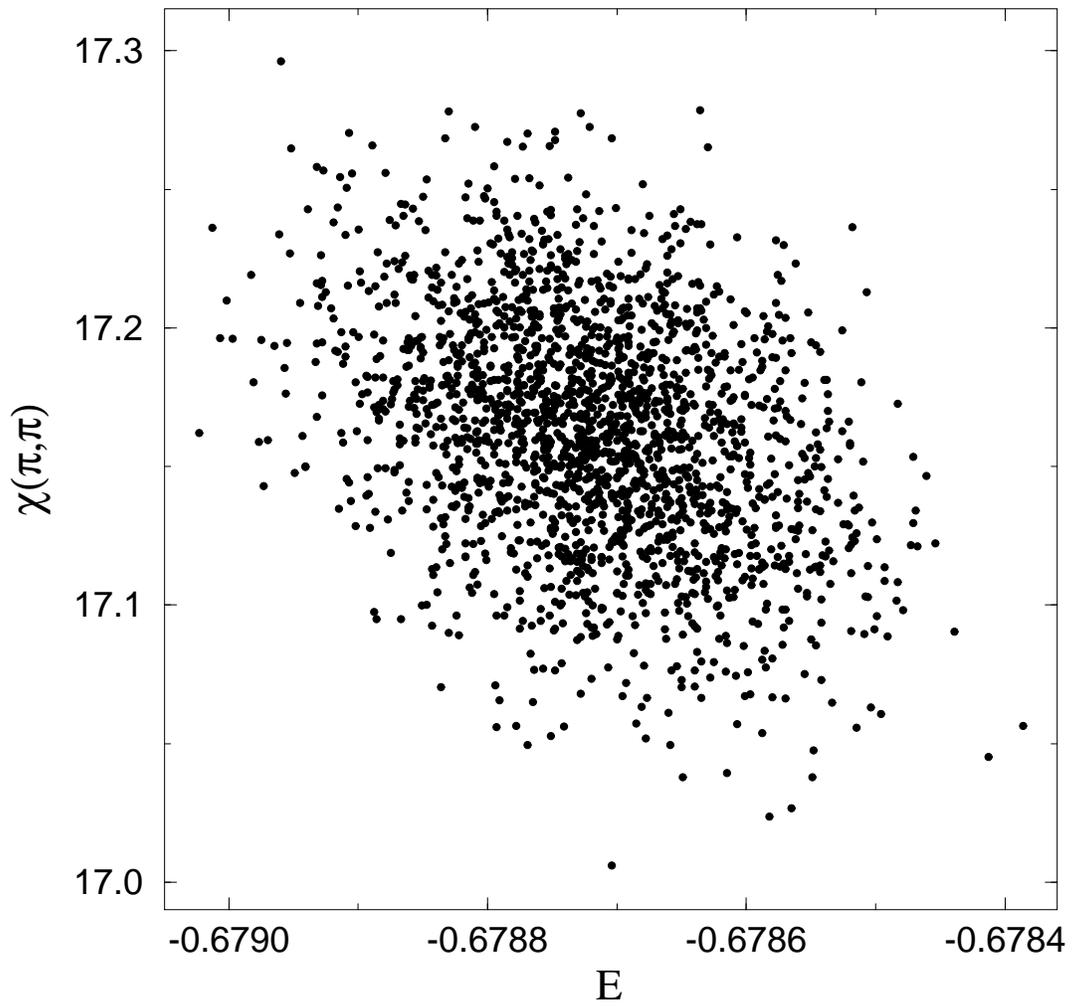
\begin{figure}
\caption{Correlations between  the staggered susceptibility and the 
energy as obtained from the nearest-neighbor correlation function 
for a $6\times 6$ lattice at $\beta=48$. Due to the very weak covariance, 
only a modest increase in accuracy can be achieved for $\chi(\pi,\pi)$.}
\label{figcovsus}
\end{figure}

Unfortunately, not all quantities exhibit a strong covariance with $C(1,0)$. 
Fig.~\ref{figcovsus} shows results for the staggered susceptibility 
(\ref{xpi}) of a $6 \times 6$ system. In this case there is only a very 
weak covariance, and hardly any gain in accuracy can be achieved.

It is easy to understand why the covariance with $C(1,0)$ is particularly 
strong for $S(\pi ,\pi)$ (or indeed any equal-time spin correlation): The 
system is rotationally invariant, but the simulation generates configurations
in a representation where the $z$-direction is singled out, and only this
component of the correlation function is measured (the other components
are not easily measurable, which is the case also with standard worldline
methods). Measurements based on a particular set of configurations (a single 
bin or a bootstrap sample) will inevitably be affected by some deviations 
from perfect rotational invariance. This is manifested as amplitude 
fluctuations in the particular spin component measured, and cause the 
covariance effects seen in the data discussed above. The ability of the
local Monte Carlo updates to rotate the direction of the antiferromagnetic
order in spin space diminishes with increasing size, leading to large
statistical fluctuations in the conventional estimate of the correlations.
The fact that the energy fluctuations do not increase with the system size 
can, in the same way, be traced to the rotationally invariant nature of the 
estimator (\ref{energy}). A somewhat more formal discussion of the covariance 
error reduction scheme can be found in Ref.~\onlinecite{covar}.

\section{Results}

Simulations of $L \times L$ systems with $L \le 16$ (only even $L$) were 
carried out at inverse temperatures $\beta =4L$ and $8L$. Within statistical 
errors the results are indistinguishable, indicating that in both cases the 
ground state completely dominates the behavior of the calculated quantities. 
This can also be checked using the finite-size singlet-triplet gap scaling 
predicted from chiral perturbation theory.\cite{hasenfratz} For $L=16$ and 
$\beta=128$, this gives an estimate of $\sim 10^{-7}$ for the relative error 
in the calculated ground state energy due to excited states (note that since 
the simulations are carried out in the canonical ensemble, only $m^z=0$ 
states are mixed in). For the smaller systems the errors are even smaller 
(the gap scales as $1/N$). All results discussed here are for $\beta=8L$.
\cite{wastenote}

The statistical errors of the calculated energies are as small as 
$\approx 10^{-5}$ for all 
$L$ studied. This accuracy exceeds by a factor 5-6 the the most accurate 
results previously reported for $L=4-16$; the GFMC results by 
Runge.\cite{runge2} Comparing the two sets of results, they agree for 
$L \le 8$, but for the larger sizes the GFMC data is consistently higher by 
$2-3$ GFMC error bars. Given the agreement to a relative accuracy of less 
than $10^{-5}$ between the SSE result for $L=6$ and the exact result, and 
the non-approximate nature of the algorithm, it is hard to see why there 
should be any systematic errors in the SSE data for the larger lattices. Note
that any remaining finite-temperature effects would lead to an overestimation 
of the energy, and hence could not explain the discrepancy with the GFMC data.
As discussed above, care has been taken to verify that in fact the finite 
temperature effects are well below the statistical errors. GFMC calculations, 
on the other hand, are in general expected to be affected by a small 
systematic error originating from ``population control'' of the varying 
number of ``random walkers'' used in that type of simulation. In Runge's 
calculation, attempts were made to remove such bias more effectively than 
in previous \cite{carlson,trivedi} GFMC calculations. However, a small 
remaining systematic error could not be ruled out, and the effect was 
expected to be an over-estimation of the energy.\cite{runge2} The 
discrepancy found here is therefore not completely surprising. The energies 
obtained with the SSE algorithm are listed in Table~\ref{tab2}. It is 
gratifying to note that the SSE energy is indeed nicely self-averaging --- 
despite the considerably fewer numbers of MC steps performed for the larger 
systems the relative errors do not differ much from the smaller sizes.

Two definitions of the sublattice magnetization have been frequently used in 
previous studies,\cite{reger} and will be used here as well. The first 
definition is in terms of the staggered structure factor,
\begin{equation}
M^2_1(L) = 3S(\pi,\pi)/L^2 ,
\label {defm1} 
\end{equation}
and the second one uses the spin-spin correlation function at the 
largest separation on the finite lattice,
\begin{equation}
M^2_2(L) =  3C(L/2,L/2). 
\label{defm2} 
\end{equation}
The factors $3$ are included to account for rotational averaging
of the $z$-component of the correlations. $M_1(L)$ and $M_2(L)$ should, of 
course, scale to the same sublattice magnetization in the thermodynamic 
limit. Using covariance error reduction, both were determined to within 
statisticals errors of $\approx 10^{-4}$ (slightly larger for the largest 
systems). This accuracy also exceeds that of previous studies. The 
results for both $S(\pi,\pi)$ and $C(L/2,L/2)$ are listed i 
Table~\ref{tab2}.

As discussed in Sec.~II, the spin stiffness can be obtained directly by
measuring the square of the fluctuating winding number. However, as
pointed out recently by Einarsson and Schulz,\cite{einarsson} the two terms 
in Eq.~(\ref{stiffdef}) have different leading size-corrections; $\sim 1/L^3$
for $E$ and $\sim 1/L$ for $\Lambda _s$. Therefore, $\Lambda _s$ is also 
calculated separately. There is a small discrepancy between the QMC 
results for $L=4$ and $L=6$, and the Lanczos results reported in 
Ref.~\onlinecite{einarsson}. Adjusting for different factors in the 
definitions, the Lanczos results are $\Lambda_s(4)=0.04832$ and 
$\Lambda_s(6)=0.06723$, whereas the QMC results obtained here are 
$\Lambda_s(4)=0.04841(2)$ and $\Lambda_s(6)=0.06791(3)$. The reason for 
the discrepancy is not clear, but carrying out $4\times 4$ exact 
diagonalizations with weak twist-fields included in the Hamiltonian, and 
subsequently calculating the derivative in Eq.~(\ref{stiffderiv}) numerically,
gives $\Lambda_s(4)=0.04840$, in good agreement with the QMC result. Hence, 
there is reason to believe that the QMC results are correct.

The uniform susceptibility has typically been obtained in numerical studies
via a definition in terms of the singlet-triplet excitation gap.
\cite{gross,runge2,schulz2} Here a different approach is taken, not requiring
simulations in the $S=1$ sector. The ${\bf q}$-dependent susceptibility,
\begin{equation}
\chi(q_x,q_y)={1\over N} \sum\limits_{j,k} {\rm e}^{i(q_xj+q_yk)} \chi (j,k),
\label{chiq}
\end{equation}
is calculated using Eq.~(\ref{susc}), and its value at the longest wavelength, 
$q_1=2\pi/L$, is taken as the definition of the finite-size uniform 
susceptibility [due to the finite-size gap and the conserved magnetization, 
$\chi (q=0)$ of course vanishes identically]. In order to give the correct 
transverse susceptibility of a system with broken symmetry in the 
thermodynamic limit, the result has to be adjusted by factor $3/2$. 
Hence, the definition is
\begin{equation}
\chi_\perp (L) = \hbox{$3 \over 2$}\chi (2\pi/L,0).
\end{equation}
The spinwave velocity can be obtained from the infinite-size values of 
$\rho_s$ and $\chi_\perp$ according to the general hydrodynamic relation
\begin{equation}
c = \sqrt{\rho_s /\chi_\perp} .
\label{hydro}
\end{equation}

The above quantities will now be scaled to the thermodynamic limit using
$\chi^2$ fits to appropriate scaling forms. Chiral perturbation theory gives 
the following scaling behavior for the ground state energy and the sublattice 
magnetization defined according to Eq.~(\ref{defm1}) [parameters without the 
argument $L$ will henceforth denote the infinite-size values]:
\begin{mathletters}
\begin{eqnarray}
E(L) & = & 
E + \beta c {1\over L^3} + {c^2 \over 4 \rho_s}{1\over L^4} + \ldots ,
\label{escale} \\
M^2_1(L) & = & 
M^2 + \alpha {M^2 \over c \chi_\perp}{1\over L} + \ldots , \label{mscale}
\end{eqnarray}
\label{emscale}
\end{mathletters}
where $\alpha = 0.62075$ and $\beta=-1.4377$.\cite{hasenfratz} The leading
corrections have been obtained also from renormalization group calculations 
for the nonlinear $\sigma$ model,\cite{neuberger,fisher} and their orders 
also agree with spinwave theory.\cite{huse}

Spinwave theory gives that the leading corrections to $\Lambda_s$ and $M_2$ 
are $\sim 1/L$,\cite{einarsson,huse} and this should be the case also for 
$\chi_\perp (L)$ due to the linear spinwave spectrum for small $q$. To the 
author's knowledge, there are no more detailed predictions for the scaling 
behavior of these quantities. 

Individually fitting all the parameters, it is found that the high accuracy
of $E(L)$ necessitates the inclusion also of a term $\sim 1/L^5$ in
Eq.~(\ref{escale}). Both $M^2_1(L)$ and $M^2_2(L)$ require corrections up
to order $1/L^3$. The QMC results for $\Lambda_s(L)$ and $\chi_\perp (L)$ 
are less accurate, and only linear and quadratic terms are needed. Hence, 
the following size dependences are assumed
\begin{mathletters}
\begin{eqnarray}
E(L)   & = & E + {e_3 \over L^3} + {e_4 \over L^4} + {e_5 \over L^5} \\
M^2_1(L) & = & M^2 + {m_1 \over L} + {m_2 \over L^2} + {m_3 \over L^3} \\
M^2_2(L) & = & M^2 + {n_1 \over L} + {n_2 \over L^2} + {n_3 \over L^3} \\
\Lambda_s(L) & = & \Lambda_s + {l_1 \over L} + {l_2 \over L^2} \\
\chi_\perp(L) & = & \chi_\perp + {x_1 \over L} + {x_2 \over L^2}.
\end{eqnarray}
\label{allscale}
\end{mathletters}
The predicted scaling forms (\ref{emscale}), together with the hydrodynamic
relation (\ref{hydro}) and the expression (\ref{stiffdef}) for the spin
stiffness $\rho_s$, imply the following constraints among the parameters and 
size-corrections:
\begin{mathletters}
\begin{eqnarray}
\Lambda_s & = & -(1/3)[E + 2 \alpha M^2 e_3 /(\beta m_1)] \label{const2} \\
\chi_\perp & = & \alpha \beta M^2 /(m_1 e_3) \label{const3} \\
e_4 &  = & m_1 e_3 /(4\alpha\beta M^2) \label{const1} .
\end{eqnarray}
\label{constraints}
\end{mathletters}
All the scaling forms (\ref{allscale}) are hence coupled to a high degree,
and a good simultaneous fit of all parameters will strongly support the
field theoretical predictions (\ref{emscale}).

Data for all sizes $L=4-16$ can be included in fits with good values of 
$\chi^2$ per degree of freedom ($\chi^2/$DOF), except in the case of 
$\chi_\perp (L)$ for which $L=4$ has to be excluded (not surprising, since 
the smallest wave-vector $q_1$ used in the definition is as large as 
$\pi /2$ for $L=4$). Both $\Lambda_s(16)$ and $\chi_\perp (16)$ have error 
bars too large to be useful, and are therefore also excluded.

Extrapolating the infinite-size parameters from fits to a small number of 
points, one has to take into account the fact that there are higher-order
corrections present, which by necessity have been neglected in the scaling 
forms used. The statistical errors of the extrapolated parameters may be 
smaller than the systematic errors introduced due to this neglect (even though
the fit may be good). In order to minimize this type of subtle errors, the 
$L=4$ data were excluded from all the fits discussed in the following.
This leads to larger statistical fluctuations but should significantly reduce 
the risk of underestimating the errors (the largest neglected correction to 
$E$ is $11$ times larger for $L=4$ than for $L=6$, and for the other 
quantities $3-5$ times larger).

Before considering the full fit (\ref{allscale}) with all the constraints
(\ref{constraints}), it is instructive to consider first the results of
individual, unconstrained fits to all the different quantities. The effects 
of including the constraints can then be judged in light of these results.

Completely independent fits give $E=-0.66943(2)$, $M=0.3062(6)$ [from 
$M_1(L)$], $M=0.3068(9)$ [from $M_2(L)$], $\rho_s = 0.179(4)$, and 
$\chi_\perp = 0.063(1)$. Note that $M_1(L)$ and $M_2(L)$ give the same 
sublattice magnetization $M$ within statistical errors, as they should.
Using (\ref{hydro}) the spinwave velocity is $c=1.69(2)$. These parameters
are in good general agreement with previous calculations, except that the 
energy is slightly lower than the best GFMC estimate,\cite{runge2} due to 
the discrepancies in the finite-size data discussed above. The sublattice 
magnetization is a bit lower than the recent result by Beard and Wiese 
\cite{beard} [$M=0.3085(2)$].

The consistency with the scaling forms (\ref{emscale}) can of course also be 
tested with these independent fits. The leading energy correction is found to
be $e_3 = -2.43(11)$, whereas Eq.~(\ref{escale}) in combination with the above
estimate for $c$ gives $e_3 = \beta c = 2.43(3)$.  The constant of the linear 
term in $M_1(L)$ is $m_1 = 0.574(9)$, whereas the right hand side of the 
scaling form (\ref{mscale}) gives $m_1 = \alpha M^2/(c\chi_\perp) = 0.550(10)$.
The subleading energy correction of the fit is $e_4 = 4(1)$, and
Eq.~(\ref{escale}) gives $e_4 = c^2 /(4 \rho_s) = 4.0(1)$. Hence, there 
is good consistency with the predicted scaling forms, within a few percent
for the leading terms, but with a statistical uncertainty as large as 
$\approx 25$\% for the subleading energy correction.

The leading corrections have been derived in several different ways.
\cite{neuberger,fisher,hasenfratz} Given then the good numerical agreement 
found above, it is now reasonable to enforce the constraints (\ref{const2})
and (\ref{const3}) which involve these terms. From such a fit, with the 
constraint on the subleading energy correction (\ref{const1}) left unenforced,
one can get a better estimate of the size of the subleading term. The 
constraint that $M_1(L)$ and $M_2(L)$ extrapolate to the same $M$ is 
also enforced. 

One of the early nonlinear $\sigma$ model calculations of the finite-size 
behavior indicated that the subleading correction to $E$ was $\sim 1/L^5$,
not $\sim 1/L^4$. \cite{fisher} Previous numerical calculations were not 
accurate enough to distinguish between these forms. However, Runge noted that 
the value of $c$ extracted from the leading correction was in rather poor 
agreement with other estimates if a $1/L^5$ subleading correction was used, 
and that a slightly better value was obtained using $1/L^4$. Even with the 
accuracy of the QMC results for $E(L)$ obtained here, individual fits using 
the two different subleading corrections (and including also the next 
higher-order correction in both cases) cannot by themselves definitely rule 
out the absence of the $1/L^4$ term, although the fit including it is better. 
However, it is not at all possible to obtain a good fit constrained by 
(\ref{const2}) and (\ref{const3}) without the $1/L^4$ term, $\chi^2/$DOF 
being as high as $\approx 50$ in this case. With the $1/L^4$ term 
$\chi^2/DOF$$\approx 0.9$. Hence, knowing the constraints on the leading 
corrections, the present data unambiguously require that the subleading 
energy term is $\sim 1/L^4$.

The parameters obtained in the partially constrained fit are: 
$E=-0.669436(5)$, $M=0.3071(3)$, $\rho_s = 0.176(2)$, 
$\chi_\perp = 0.0623(10)$, and $c= 1.681(14)$. 
The statistical errors are here significantly reduced relative
to the previous unconstrained fits, and the two sets of parameters are 
consistent with each other. The leading corrections to $E$ and $M$ are now, 
of course, in complete agreement with the theoretical prediction. The 
subleading energy correction of the fit is $e_4 = 4.17(23)$, whereas 
Eq.~(\ref{escale}) with the above parameters gives $e_4 = c^2 /(4 \rho_s) = 
4.01(7)$. Hence, it is now also confirmed, at the $5$\% accuracy level, 
that the size of the subleading energy correction agrees with the chiral 
perturbation theory prediction by Hasenfratz and Niedermayer.\cite{hasenfratz}
It is remarkable that the derivation of this very detailed theoretical result
is based purely on symmetry and dimensionality considerations.\cite{hasenfratz}

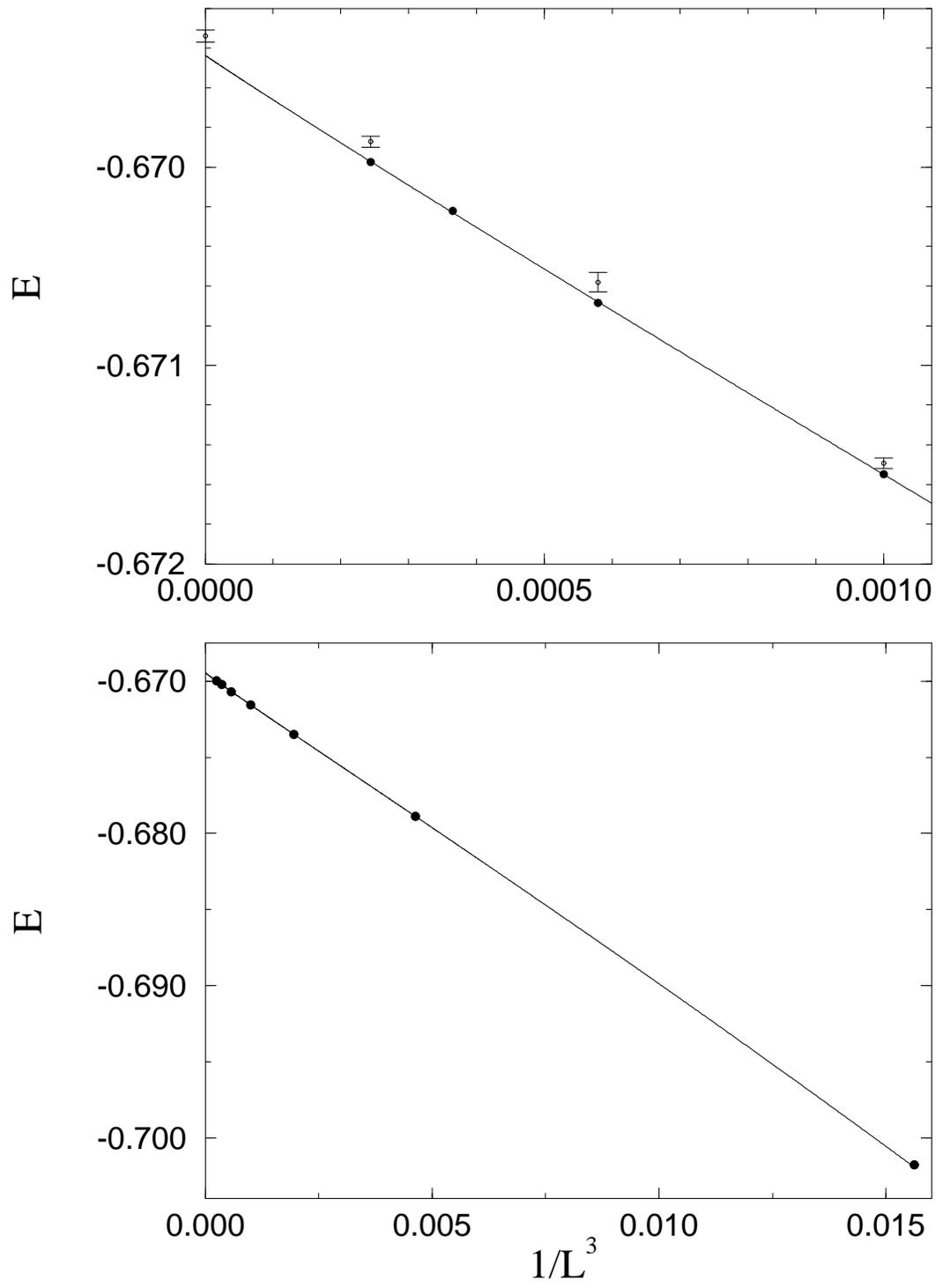
\begin{figure}
\caption{The ground state energy vs.~$1/L^3$ on two different scales.
The solid circles are the QMC data. On the scale of the top graph, the
error bars are smaller than half the radius of the circles. The smaller
circles with error bars in the top graph are the GFMC and extrapolated results 
by Runge.\protect{\cite{runge2}} The result of the constrained fit
including the $L > 4$ data is indicated by the curves.}
\label{figenergy}
\end{figure}

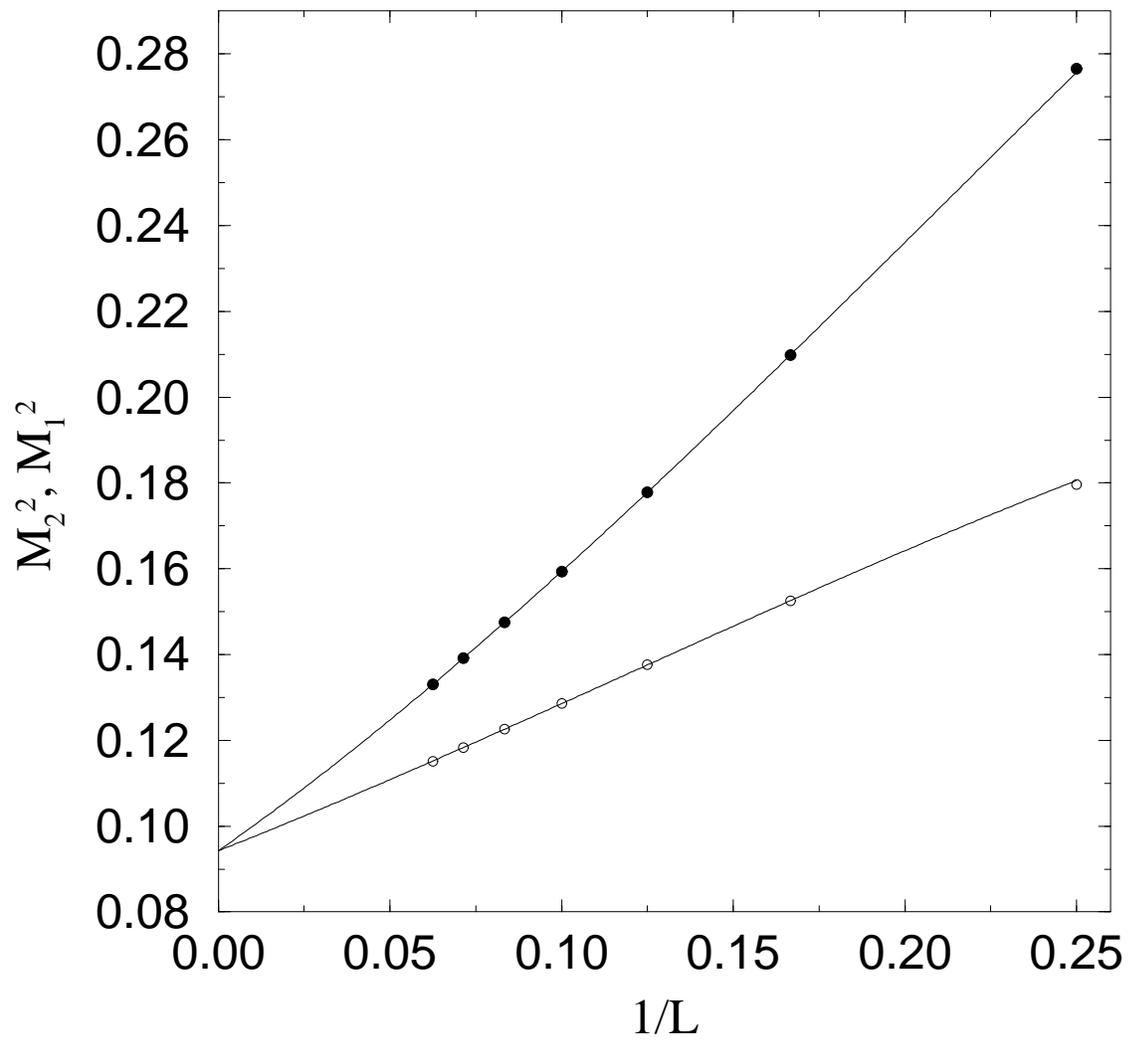
\begin{figure}
\caption{The sublattice magnetization squared, as defined by 
Eqs.~(\protect{\ref{defm1}}) and (\protect{\ref{defm2}}), vs.~inverse
system size. The solid and open circles are the QMC data for $M_1(L)$ 
and $M_2(L)$, respectively, and the curves are the results of the constrained
fit including the $L > 4$ data. The QMC error bars are much smaller than 
the circles.}
\label{figmag}
\end{figure}

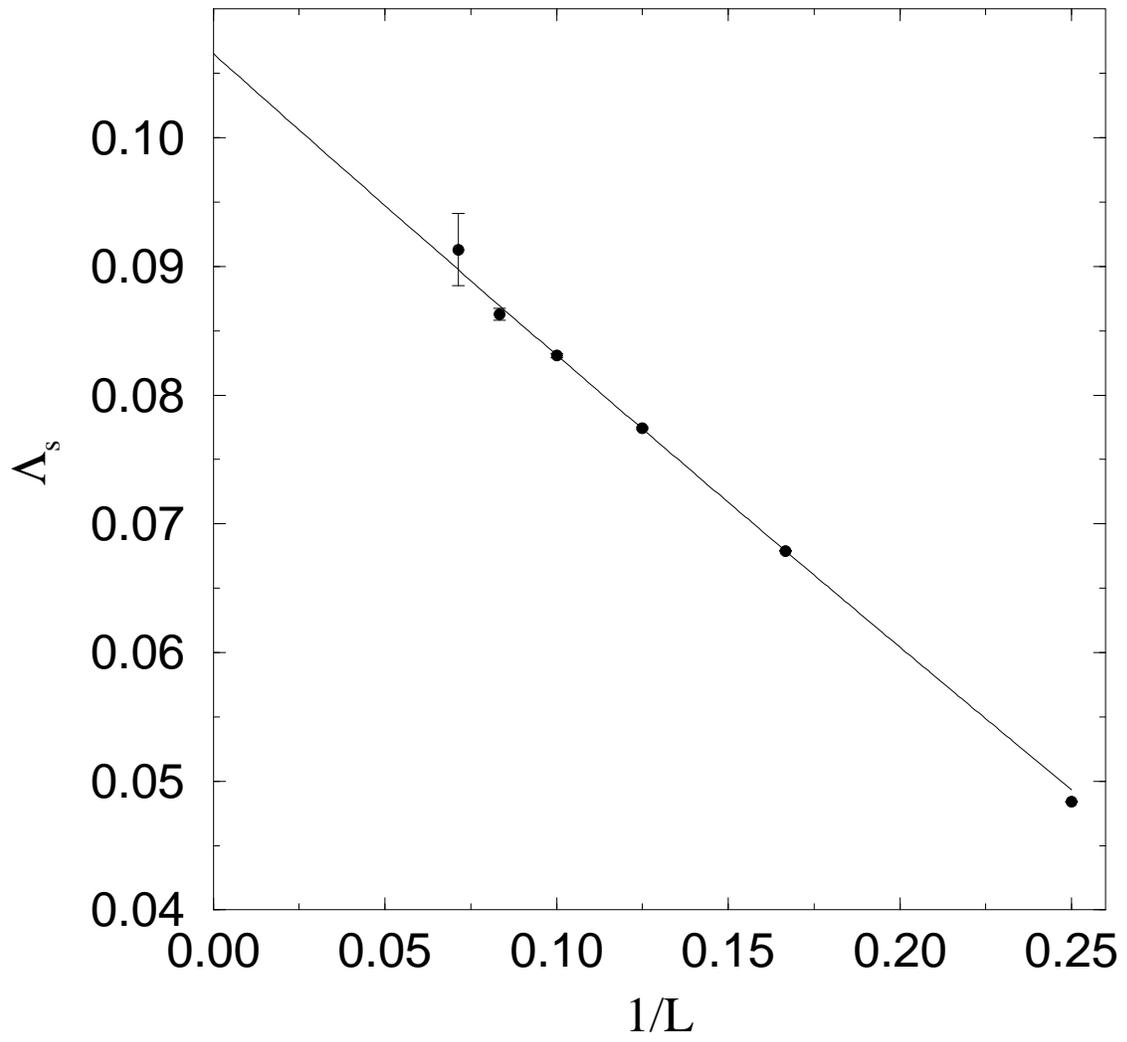
\begin{figure}
\caption{The spin current-current correlator, Eq.~(\protect{\ref{lambdas}}),
vs.~the inverse system size, along with the result of the coupled fit
to the $L > 4$ data.}
\label{figlambda}
\end{figure}

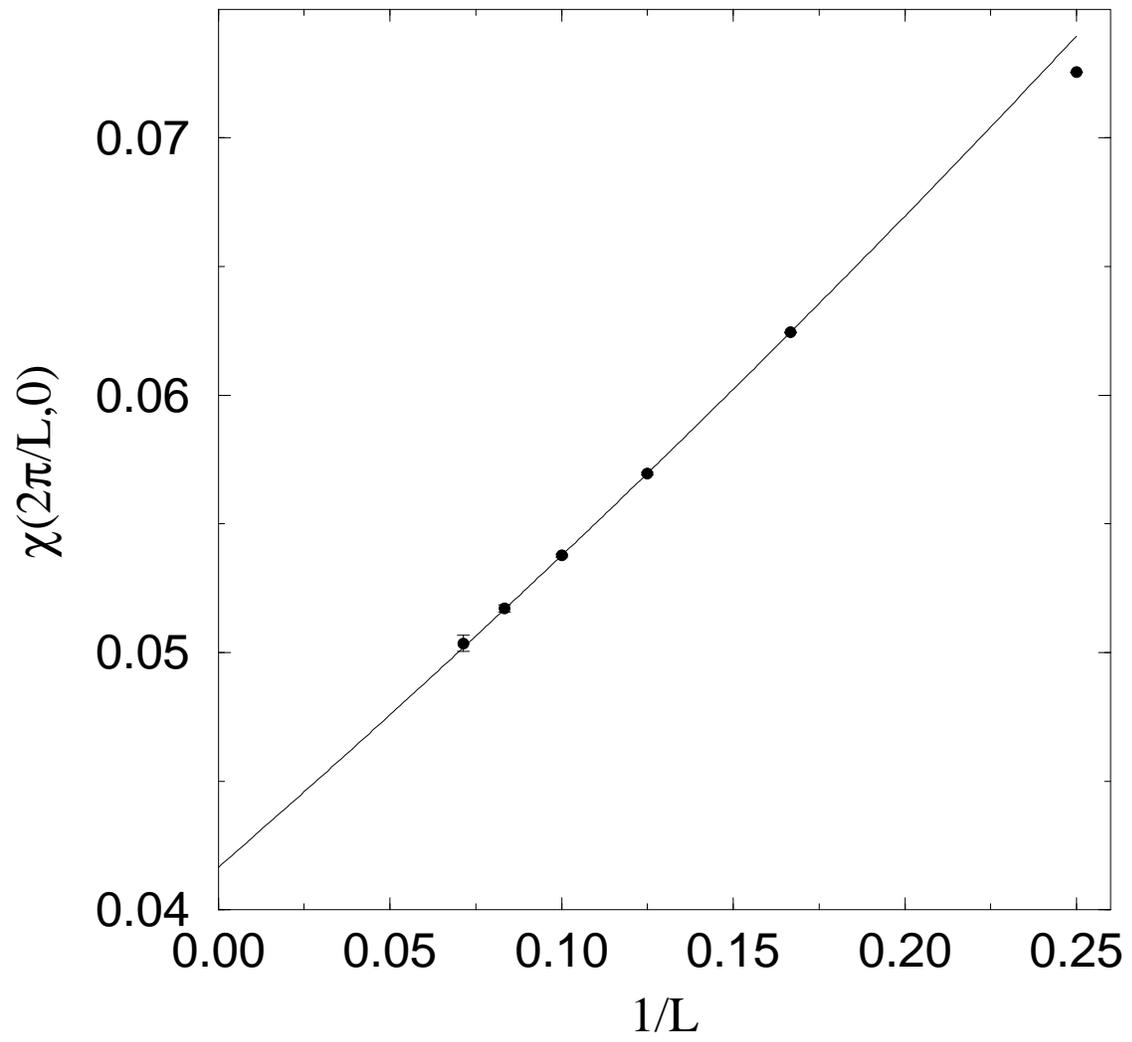
\begin{figure}
\caption{The long-wavelength spin susceptibility [$2/3$ of $\chi_\perp (L)$]
vs.~inverse system size, along with the result of the constrained fit to the 
$L > 4$ data.}
\label{figsusc}
\end{figure}

With the subleading energy correction now firmly established, the last 
constraint (\ref{const1}) can also be enforced, and the parameters of 
this fit are taken as the final results. This coupled fit still has 
$\chi^2/$DOF$\approx 0.9$ (the total number of parameters is $14$, and a 
total of $28$ data points were used). Looking at $\chi ^2$ for the five 
individual curves, they all represent good fits to their respective data 
sets. Hence, the constrained fit is in all respects a good one. All the 
ground state parameters obtained from the fit are listed in Table~\ref{tab3},
along with the leading and subleading corrections to the energy and the 
sublattice magnetization. There are only minor changes relative to the 
previous partially constrained fit, the main improvement in accuracy being 
for the spinwave velocity. The errors of the parameters were calculated 
using the bootstrap method,\cite{bootstrap} i.e., fits were carried out for 
a large number of bootstrap samples of the QMC data, and the error is defined
as one standard deviation of the parameters of those fits. As explained in 
Sec.~II-B, this should be a very accurate method for calculating statistical
errors even in highly nonlinear situations such as the constrained $\chi ^2$ 
fit. The QMC data used, along with the fitted curves, are shown in 
Figs.~\ref{figenergy}--\ref{figsusc}. 

In the figures, it can be observed that even though the $L=4$ data are not 
included in the fit, the fitted curves extrapolated to $L=4$ are quite close
to these QMC points, except in the case of $\chi_\perp (4)$ (for which this 
point cannot even be included in an individual fit). In fact, calculating 
also the statistical errors of the extrapolations to $L=4$ gives strong 
further support to the reliability of the procedures used. For all quantities 
except the energy, the statistical errors are found to be comparable at $L=4$ 
and $L=\infty$. For the energy the fluctuations  are more than $20$ times 
larger at $L=4$, due to the high order of the leading correction. Both $E(4)$ 
and $\Lambda_s(4)$ are within one standard deviation of the extrapolated 
results. The sublattice magnetizations $M_1(4)$ and $M_2(4)$ both deviate 
by $2.5$ standard deviations, and $\chi_\perp$ deviates by $3$ standard 
deviations. These rather small deviations clearly indicate that there are 
only minor effects of neglected higher-order corrections. The extrapolations 
to $L = \infty$ should of course be significantly less sensitive to systematic
errors, since the neglected corrections rapidly vanish for the larger sizes. 
As already discussed above, the neglected corrections are several times larger
at $L=4$ than at $L=6$ (the smallest size considered in the fits). Hence, any 
remaining systematic errors in the infinite-size extrapolations listed in 
Table~\ref{tab3} should be well below the indicated statistical errors.

\section{Summary and Discussion}

An extensive study of the ground state parameters of the 2D Heisenberg model
has been presented. Using the  Stochastic Series Expansion QMC method
\cite{sse1,sse2} in combination with a data analysis scheme utilizing 
covariance effects,\cite{covar} results of unprecedented accuracy were 
obtained for the ground state energy and the sublattice magnetization for 
systems of linear dimensions up to $L=16$. The long-wavelength susceptibility
and the spin stiffness were also directly calculated in the simulations.

The QMC data was extrapolated to the thermodynamic limit using scaling forms 
predicted from chiral perturbation theory,\cite{hasenfratz} supplemented by 
higher-order terms necessary to obtain good fits. Both the leading and 
subleading corrections were found to agree in magnitude with the theoretical 
predictions to within a few percent. This is the first numerical verification 
of the predictions of chiral perturbation theory \cite{hasenfratz} 
to subleading order.

The ground state energy extracted from the fit is the most accurate estimate
obtained to date, and is slightly higher than the best Green's function Monte 
Carlo result.\cite{runge2} This discrepancy is most likely due to a 
``population control'' bias in the GFMC calculation.\cite{runge2} The spin 
stiffness and the sublattice magnetization are both lower than the results 
of a recent low-temperature loop algorithm QMC study.\cite{beard} The 
discrepancy appears to be marginally larger than what could be explained 
by statistical fluctuations alone. For the spinwave velocity the results 
are consistent with each other.

In the QMC study by Beard and Wiese \cite{beard} the size and temperature
dependence of the uniform and staggered susceptibilities were fit to scaling 
forms from chiral perturbation theory. Hence, the underlying theory for 
analyzing the data is the same as used in the present study, but the physical 
quantities used are different, as is the temperature regime (low but finite
$T$ versus $T=0$ in this study). Since both QMC algorithms are ``exact'', the 
discrepancies must originate from the scaling procedures. In this paper the 
effects of neglected higher-order corrections were discussed, and attempts 
were made to minimize these as much as possible. Furthermore, as a 
quantitative check of remaining effects of this nature, the calculated 
scaling functions were extrapolated to lattices {\it smaller} than 
the smallest size used in the fit ($L=6$). The close ageement with the actual 
calculated results, along with the high orders of the largest neglected 
corrections, show that any effects of the higher-order terms on the 
extrapolations to infinite size should be well below the carefully
computed statistical errors. It can be noted that Beard and Wiese also 
included subleading corrections,\cite{beard} but not to the same high orders 
as was necessary in the present study (due to the high accuracy of the SSE
results for the energy and the sublattice magnetization). Another reason
to believe that the present study is more reliable is that the fit involves 
in a direct manner the $T=0$ finite-size definitions of the same infinite-size
parameters sought, not functions of those parameters. 

In combination with the covariance error reduction scheme,\cite{covar}
the SSE method is a very efficient method for calculating correlation 
functions of isotropic spin models, as exemplified by the results presented 
here. The covariance scheme is most efficient in cases where there are strong 
long-ranged correlations (where the ``bare'' estimator for the correlation 
function does not behave well). It is currently being applied in a study the 
temperature dependence of the correlation length of the weakly coupled 
Heisenberg bilayer, for larger lattices and lower temperatures than previously 
\cite{bilayer1} possible. This is motivated by recent results obtained from 
a mapping to a nonlinear $\sigma$ model for this system,\cite{yin}
predicting a much faster divergence of the correlation length as $T\to 0$
than for the single layer. The SSE algorithm has also proven useful in the 
case of critical, or near-critical systems, such as the bilayer Heisenberg 
model close to its quantum critical point. More accurate finite-size scalings
than previously reported for this,\cite{bilayer2} as well as other models 
exhibiting quantum critical behavior,\cite{dimrand} are possible with the 
data enhanced using the covariance method.

QMC methods based on the loop-cluster algorithm,\cite{evertz} have proven
to be very efficient in several studies of $S=1/2$ Heisenberg models,
\cite{wiese,beard,loopstudies} and are clearly more efficient than the SSE 
method in many cases. For example, it was shown here that the covariance 
method cannot significantly improve the accuracy of calculated static 
susceptibilities, which appear to be very accurately given in loop algorithm 
simulations.\cite{beard} The method used here for calculating the spin
stiffness directly from the winding number fluctuations would probably also
be more accurate with loop algorithms. However, it is not clear whether 
loop algorithms can easily produce more accurate results for equal-time 
correlation functions or energies than those presented in this paper. 

The methods discussed here can also be easily extended to higher-spin models. 
In fact, the covariance scheme has an additional advantage for $S > 1/2$, 
in that the on-site correlation $(S^z_i)^2$ is known exactly, but fluctuates 
in the simulation and exhibits strong covariance with other correlation
functions.\cite{covar} Detailed studies of various higher-spin models
should therefore now be feasible also in $2D$.

\section{Acknowledgments}

I would like to thank B. Beard, D. Scalapino, R. Sugar, and U.-J. Wiese for 
discussions. This work was supported by the National Science Foundation under 
Grant No.~DMR-89-20538.

\vfill\eject

\begin{table}
\begin{tabular}{ldd}
 ~                 &   ~~~~$L=4$     &   ~~~~$L= 6$      \\ \hline
 $E_{\rm exact}$   &   -0.701780     &    -0.678872      \\
 $E_1$             &   -0.701777(7)  &    -0.678873(4)  \\
 $E_2$             &   -0.70177(6)   &    -0.67872(9)     \\
 $S_{\rm exact}$   &   1.47481       &    2.5180        \\ 
 $S_1$             &   1.47480(4)    &    2.51799(6)    \\
 $S_2$             &   1.4747(2)     &    2.5168(8)     \\
\end{tabular}
\vskip2mm
\caption{Comparisons of QMC and exact results for $4\times 4$ and
$6\times 6$ lattices. $E_{\rm exact}$ is the exact result for the
ground state energy per spin, $E_1$ is the QMC estimate obtained from the
average length of the series expansion according to
Eq.~(\protect{\ref{energy}}), and $E_2$ is the estimate obtained from
the nearest-neighbor correlation function $C(1,0)$ calculated according
to Eq.~(\protect{\ref{correl}}). $S_{\rm exact}$ is the exact staggered
structure factor, $S_1$ is the QMC result with the accuracy increased
by using the covariance with $E_2$, and $S_2$ is the estimate using directly
the sum of the correlation functions, Eq.~(\protect{\ref{spi}}).
The exact $L=6$ results are from Ref.~\protect{\onlinecite{schulz2}}.}
\label{tab1}
\end{table}

\begin{table}
\begin{tabular}{lddd}
 $L$ &  ~~~~$-E$  & ~~~~~~$S(\pi,\pi)$  & ~~~~~~~~~~~$C(L/2,L/2)$   \\ \hline
  4  &   0.701777(7)  &  1.47480(4)   &  0.059872(5)     \\
  6  &   0.678873(4)  &  2.51799(6)   &  0.050856(3)     \\
  8  &   0.673487(4)  &  3.7939(2)    &  0.045867(5)     \\
 10  &   0.671549(4)  &  5.3124(3)    &  0.042851(6)     \\
 12  &   0.670685(5)  &  7.0780(7)    &  0.040873(9)     \\
 14  &   0.670222(7)  &  9.090(1)     &  0.03945(1)      \\
 16  &   0.669976(7)  & 11.352(2)     &  0.03839(2)      \\
\end{tabular}
\vskip2mm
\caption{QMC results for the ground state energy, the staggered structure 
factor, and the spin-spin correlation function at distance ${\bf r}=(L/2,L/2)$,
The accuracies of the results for $S(\pi,\pi)$ and $C(L/2,L/2)$ were
increased by using covariance effects, as discussed in Sec.~II. The 
simulations were carried out at inverse temperatures $\beta = 8L$, and
the total number of MC steps performed for the different systems were
$2.5 \times 10^8$ ($L=4$), $3 \times 10^8$ ($L=6$), $10^8$ ($L=8$),
$10^8$ ($L=10$), $3\times 10^7$ ($L=12$), $1.5 \times 10^7$ ($L=14$),
and $10^7$ ($L=16$).}
\label{tab2}
\end{table}

\begin{table}
\begin{tabular}{ldd}
 parameters        &   ~  &     \\ \hline
 $E$               &   -0.669437(5)     &     \\
 $M$               &    0.3070(3)       &     \\
 $\rho_s$          &    0.175(2)        &     \\
 $\chi_\perp$      &    0.0625(9)       &     \\ 
 $c$               &    1.673(7)        &     \\ \hline
 size corrections  &   ~  &     \\ \hline
 $e_3$             &   -2.405(10)       &     \\
 $e_4$             &    4.00(6)         &     \\
 $m_1$             &    0.560(6)        &     \\
 $m_2$             &    1.08(5)         &     \\
\end{tabular}
\vskip2mm
\caption{The ground state parameters and the leading and subleading 
corrections to $E$ and $M$, as obtained from a fit fully constrained
by the predictions of chiral perturbation theory.}
\label{tab3}
\end{table}

\begin{references}

\bibitem{csh} 
S. Chakravarty, B. I. Halperin, and D. R. Nelson, Phys. Rev. Lett. {\bf 60},
1057 (1988); Phys. Rev. B {\bf 39}, 2344 (1989).

\bibitem{s1order}
E. J. Neves and J. F. Peres, Phys. Lett. {\bf 114A}, 331 (1986);
F. J. Dyson, E. H. Lieb, and B. Simon, J. Stat. Phys. {\bf 18}, 335 (1987);
I. Affleck, T. Kennedy, E. H. Lieb, and H. Tasaki, Commun. Math. Phys. 
{\bf 155}, 477 (1988).

\bibitem{reger}
J. D. Reger and A. P. Young, Phys. Rev. B {\bf 37}, 5978 (1988).

\bibitem{spinwave1} 
P. W. Anderson, Phys. Rev. {\bf 86}, 694 (1952).

\bibitem{spinwave2} T. Oguchi, Phys. Rev. {\bf 117}, 117 (1960).

\bibitem{singh}
R. R. P. Singh, Phys. Rev. B {\bf 39}, 9760 (1989); R. R. P. Singh and
D. A. Huse, {\it ibid}, {\bf 40}, 7247 (1989).

\bibitem{hamer}
C. J. Hamer, Z. Weihong, and P. Arndt, Phys. Rev. B {\bf 46}, 6276 (1992).

\bibitem{igarashi}
J. Igarashi, Phys. Rev. B {\bf 46}, 10763 (1992).

\bibitem{canali}
C. M. Canali and M. Wallin, Phys. Rev. B {\bf 48}, 3264 (1993).

\bibitem{barnes}
T. Barnes and E. S. Swanson, Phys. Rev. B {\bf 37}, 9405 (1988).

\bibitem{carlson}
J. Carlson, Phys. Rev. B {\bf 40}, 846 (1989).

\bibitem{gross}
M. Gross, E. S\'anches-Velasco, and E. Siggia, Phys. Rev. B {\bf 39}, 2484
(1989); {\bf 40}, 11328 (1989);

\bibitem{trivedi}
N. Trivedi and D. M. Ceperley, Phys. Rev. B {\bf 40}, 2737 (1990);
{\bf 41}, 4552 (1990).

\bibitem{liang}
S. Liang, Phys. Rev. B {\bf 42}, 6555 (1990).

\bibitem{runge1}
K. J. Runge, Phys, Rev. B {\bf 45}, 7229 (1992).

\bibitem{runge2}
K. J. Runge, Phys, Rev. B {\bf 45}, 12292 (1992).

\bibitem{sauer}
R. A. Sauerwein and M. J. de Oliveira, Phys. Rev. B {\bf 49}, 5983 (1994).

\bibitem{wiese}
U.-J. Wiese and H.-P. Ying, Z. Phys. B {\bf 93}, 147 (1994).

\bibitem{beard}
B. B. Beard and U.-J. Wiese, Phys. Rev. Lett. {\bf 77}, 5130 (1996).

\bibitem{schulz1}
H. J. Schulz, T. A. L. Ziman, Europhys. Lett. {\bf 18}, 355 (1992).

\bibitem{schulz2}
H. J. Schulz, T. A. L. Ziman, and D. Poilblanc,  
J. Physique I  {\bf 6}, 675 (1996).

\bibitem{einarsson}
T. Einarsson and H. J. Schulz, Phys. Rev. B {\bf 51}, 6151 (1995).

\bibitem{weihong}
Z. Weihong, J. Oitmaa, and C. J. Hamer, Phys. Rev. B {\bf 43}, 8321 (1991).

\bibitem{critical} 
A. Chubukov, S. Sachdev, and J. Ye, Phys. Rev. B {\bf 49}, 11919 (1994).

\bibitem{neuberger} 
H. Neuberger and T. Ziman, Phys. Rev. B {\bf 39}, 2608 (1989).

\bibitem{fisher} 
D. S. Fisher, Phys. Rev. B {\bf 39}, 11783 (1989).

\bibitem{hasenfratz}
P. Hasenfratz and F. Niedermayer, Z. Phys. B {\bf 92}, 91 (1993).

\bibitem{lancscaling}
I. Affleck, D. Gepner, H. J. Schulz and T. Ziman, J. Phys. A {\bf 22},
511 (1989); F. C. Alcaraz and A. Moreo, Phys. Rev. B {\bf 46}, 2896
(1992); S. Eggert, Phys. Rev. B {\bf 54}, R9612 (1996).

\bibitem{dmrgscaling}
K. A. Hallberg, P. Horsch, and G. Martinez, Phys. Rev. B {\bf 52},
R719 (1995); K. Hallberg, X. Q. G. Wang, P. Horsch, and A. Moreo, Phys. 
Rev. Lett. {\bf 76}, 4955 (1996).

\bibitem{dmrg}
S. R. White, Phys. Rev. Lett. {\bf 69}, 2863 (1992).

\bibitem{evertz}
H. G. Evertz, G. Lana, and M. Marcu, Phys. Rev. Lett. {\bf 70},
875 (1993).

\bibitem{suzuki1}
M. Suzuki, Prog. Theor. Phys. {\bf 56}, 1454 (1976).

\bibitem{suzuki2}
M. Suzuki, S. Miyashita, and A. Kuroda, Prog. Theor. Phys. 
{\bf 58}, 1377 (1977); M. Barma and B. S. Shastry, Phys. Rev. B {\bf 18}, 
3351 (1977); J. J. Cullen and D. P. Landau, Phys. Rev. B {\bf 27}, 297 
(1983).

\bibitem{worldline}
J. E. Hirsch, R. L. Sugar, D. J. Scalapino and R. Blankenbecler,
Phys. Rev. B {\bf 26}, 5033 (1982).

\bibitem{sse1}
A. W. Sandvik and J. Kurkij\"arvi, Phys. Rev. B {\bf 43}, 5950 (1991).

\bibitem{sse2}
A. W. Sandvik, J. Phys. A {\bf 25}, 3667 (1992).

\bibitem{covar}
A. W. Sandvik, Phys. Rev. B {\bf 54}, 14910 (1996).

\bibitem{chapter}
A. W. Sandvik, in {\it Numerical Methods for Lattice Many-Body Models},
ed. by D. J. Scalapino (to be published).

\bibitem{bilayer1}
A. W. Sandvik and D. J. Scalapino, Phys. Rev. Lett. {\bf 72}, 2777 (1994);
A. W. Sandvik, A. V. Chubukov, and S. Sachdev, Phys. Rev. B {\bf 51},
16483 (1995).

\bibitem{bilayer2}
A. W. Sandvik and D. J. Scalapino, Phys. Rev. B {\bf 53}, R526 (1996).

\bibitem{dimrand}
A. W. Sandvik and M. Veki\'c, Phys. Rev. Lett. {\bf 74}, 1226 (1995);
J. Low. Temp. Phys. {\bf 99}, 367 (1995).

\bibitem{chains}
O. A. Starykh, A. W. Sandvik, and R. R. P. Singh, Phys. Rev. B {\bf 55},
14953 (1997).

\bibitem{ssefermions}
A. W. Sandvik, D. J. Scalapino, and P. Henelius, Phys. Rev. B {\bf 50}, 10474 
(1994); A. W. Sandvik and A. Sudb{\o}, Phys. Rev. B {\bf 54}, R3746 (1996); 
Europhys. Lett. {\bf 36}, 443 (1996).

\bibitem{handscomb}
D. C. Handscomb, Proc. Cambridge Philos. Soc. {\bf 58}, 594 (1962);
{\bf 60}, 115 (1964); J. W. Lyklema, Phys. Rev. Lett. {\bf 49}, 88 (1982).

\bibitem{lee}
D. H. Lee, J. D. Joannopoulos, and J. W. Negele, Phys. Rev. B {\bf 30}, 
1599 (1984); E. Manousakis and R. Salvador, Phys. Rev. B {\bf 39},
575 (1989).

\bibitem{metropolis} 
N. Metropolis, A. Rosenbluth, M. Rosenbluth, A. H. Teller, and E. Teller,
J. Chem. Phys. {\bf 21}, 1087 (1953).

\bibitem{lieb}
E. H. Lieb and D. C. Mattis, J. Math. Phys. {\bf 3}, 749 (1962).

\bibitem{patrik}
P. Henelius {\it et al.} (unpublished).

\bibitem{scalapino}
D. J. Scalapino, S. R. White, and S. C. Zhang, Phys. Rev. B {\bf 47},
7995 (1993).

\bibitem{pollock}
E. L. Pollock and D. M. Ceperley, Phys. Rev. B {\bf 36}, 8343 (1987).

\bibitem{bootstrap}
B. Efron and G. Gong, Am. Stat. {\bf 37}, 36 (1983).

\bibitem{wastenote}
It should be noted that at low temperatures the effective amount of data 
contained in a single QMC configuration is proportional to $\beta$, since the 
finite-size gap implies an exponential decay of correlations in the ``time'' 
dimension. Hence, using temperatures lower than necessary (as is the case 
for the smaller systems studied here) does not imply a waste of resources.

\bibitem{huse}
D. A. Huse, Phys. Rev. B {\bf 37}, 2380 (1988).

\bibitem{yin}
L. Yin and S. Chakravarty, preprint (cond-mat/9703138).

\bibitem{loopstudies}
M. Troyer, H. Kontani, and K. Ueda, Phys. Rev. Lett. {\bf 76}, 3822 (1996).
M. Greven, R. J. Birgeneau, and U. J. Wiese, Phys. Rev. Lett. {\bf 77}, 
1865 (1996); B. Frischmut, B. Ammon, and M. Troyer, Phys. Rev. B {\bf 54},
R3714 (1996). M. Troyer, Zhitomirsky, and K. Ueda, Phys. Rev. B {\bf 55},
R6117 (1997).

\end{references}
\end{document}